\documentclass[iop,letter]{emulateapj}

\usepackage{natbib}
\usepackage{graphicx}
\usepackage{times}
\usepackage{apjfonts}
\usepackage{amssymb,amsmath}
\usepackage{bm} 
\usepackage{color}

\def\ltsima{$\; \buildrel < \over \sim \;$}
\def\simlt{\lower.5ex\hbox{\ltsima}}
\def\gtsima{$\; \buildrel > \over \sim \;$}
\def\simgt{\lower.5ex\hbox{\gtsima}}
\def\simless{\mathbin{\lower 3pt\hbox
   {$\rlap{\raise 5pt\hbox{$\char'074$}}\mathchar"7218$}}}   
\def\simgreat{\mathbin{\lower 3pt\hbox
   {$\rlap{\raise 5pt\hbox{$\char'076$}}\mathchar"7218$}}}   

\def\chandra{{\it Chandra}}
\def\spitzer{{\it Spitzer}}
\def\3p6{3.6$\mu$m}
\def\4p5{4.5$\mu$m}

\def\SPT2106{SPT-CL~J2106$-$5844}
\def\J0102{ACT-CL~J0102$-$4915}
\def\rxj{RXJ0152$-$1357}
\def\bullet{1E~0657$-$558}
\def\ms1054{MS~1054$-$0321}
\def\kms{km~s$^{-1}$}
\def\lcdm{$\Lambda$CDM}

\slugcomment{Draft 0.1}
\submitted{Submitted to the Astrophysical Journal, 2011 September 4;
  accepted 2012 January 4}

\shorttitle{ACT-CL~J0102$-$4915, ``El Gordo''}
\shortauthors{Menanteau et al.}

\begin{document}

\title{The Atacama Cosmology Telescope: ACT-CL~J0102$-$4915 ``El Gordo,'' a Massive Merging Cluster at Redshift 0.87}

\author{Felipe Menanteau\altaffilmark{1}\footnotemark[\dag],
John~P.~Hughes\altaffilmark{1},
Crist\'obal~Sif\'on\altaffilmark{2},
Matt~Hilton\altaffilmark{3},
Jorge~Gonz\'alez\altaffilmark{2},
Leopoldo~Infante\altaffilmark{2},
L.~Felipe~Barrientos\altaffilmark{2},
Andrew~J.~Baker\altaffilmark{1},
John~R.~Bond\altaffilmark{4},
Sudeep~Das\altaffilmark{5,6,7},
Mark~J.~Devlin\altaffilmark{8},
Joanna~Dunkley\altaffilmark{9},
Amir~Hajian\altaffilmark{4},
Adam~D.~Hincks\altaffilmark{7},
Arthur~Kosowsky\altaffilmark{10},
Danica~Marsden\altaffilmark{8},
Tobias~A.~Marriage\altaffilmark{11},
Kavilan~Moodley\altaffilmark{12},
Michael~D.~Niemack\altaffilmark{13},
Michael~R.~Nolta\altaffilmark{4},
Lyman~A.~Page\altaffilmark{7},
Erik~D.~Reese\altaffilmark{8},
Neelima Sehgal\altaffilmark{14},
Jon~Sievers\altaffilmark{4},
David~N.~Spergel\altaffilmark{6},
Suzanne~T.~Staggs\altaffilmark{7},
Edward~Wollack\altaffilmark{15}
}

\affil{$^1$Rutgers University, Department of Physics \& Astronomy, 136 Frelinghuysen Rd, Piscataway, NJ 08854, USA }
\affil{$^2$Departamento de Astronom{\'{i}}a y Astrof{\'{i}}sica,
  Facultad de F{\'{i}}sica, Pontific\'{i}a Universidad Cat\'{o}lica de
  Chile, Casilla 306, Santiago 22, Chile}
\affil{$^3$School of Physics and Astronomy, University of Nottingham, University Park, Nottingham, NG7 2RD, UK}
\affil{$^4$Canadian Institute for Theoretical Astrophysics,  University of Toronto, Toronto, ON, Canada M5S 3H8}
\affil{$^5$Berkeley Center for Cosmological Physics, LBL and
  Department of Physics, University of California, Berkeley, CA 94720, USA}
\affil{$^6$Department of Astrophysical Sciences, Peyton Hall, Princeton University, Princeton, NJ, 08544, USA}
\affil{$^7$Joseph Henry Laboratories of Physics, Jadwin Hall, Princeton University, Princeton, NJ, 08544, USA}
\affil{$^8$University of Pennsylvania, Physics and Astronomy, 209 South 33rd Street, Philadelphia, PA 19104, USA}
\affil{$^9$Department of Astrophysics, Oxford University, Oxford, OX1 3RH, UK }
\affil{$^{10}$University of Pittsburgh, Physics \& Astronomy Department, 100 Allen Hall, 3941 O'Hara Street, Pittsburgh, PA 15260, USA}
\affil{$^{11}$Department of Physics and Astronomy, The Johns Hopkins University, Baltimore, Maryland 21218-2686, USA}
\affil{$^{12}$University of KwaZulu-Natal, Astrophysics \& Cosmology Research Unit, School of Mathematical Sciences, Durban, 4041, South Africa.}
\affil{$^{13}$NIST Quantum Devices Group, 325 Broadway Mailcode 817.03, Boulder, CO, USA 80305}
\affil{$^{14}$KIPAC, Stanford University, Stanford, California, 94305, USA}
\affil{$^{15}$Code 553/665, NASA/Goddard Space Flight Center, Greenbelt, Maryland, 20771, USA}

\footnotetext[\dag]{Based on observations made with ESO Telescopes at
  the Paranal Observatory under programme ID 086.A-0425.}

\begin{abstract}

  We present a detailed analysis from new multi-wavelength
  observations of the exceptional galaxy cluster \J0102, likely the
  most massive, hottest, most X-ray luminous and brightest
  Sunyaev-Zel'dovich (SZ) effect cluster known at redshifts greater
  than 0.6. The Atacama Cosmology Telescope (ACT) collaboration
  discovered \J0102 as the most significant Sunyaev-Zeldovich (SZ)
  decrement in a sky survey area of 755 square degrees. Our VLT/FORS2
  spectra of 89 member galaxies yield a cluster redshift, $z=0.870$,
  and velocity dispersion, $\sigma_{\rm gal}=1321\pm106$~\kms. Our
  \chandra\ observations reveal a hot and X-ray luminous system with
  an integrated temperature of $T_X=14.5\pm1.0$~keV and 0.5--2.0 keV
  band luminosity of
  $L_X=(2.19\pm0.11)\times10^{45}\,h_{70}^{-2}$erg~s$^{-1}$. We obtain
  several statistically consistent cluster mass estimates; using
  empirical mass scaling relations with velocity dispersion, X-ray
  $Y_X$, and integrated SZ distortion, we estimate a cluster mass of
  $M_{200a}=(2.16\pm0.32)\times10^{15}\,h_{70}^{-1}M_\odot$.  We
  constrain the stellar content of the cluster to be less than 1\% of
  the total mass, using \spitzer\ IRAC and optical imaging.  The
  \chandra\ and VLT/FORS2 optical data also reveal that \J0102\ is
  undergoing a major merger between components with a mass ratio of
  approximately 2 to 1.  The X-ray data show significant temperature
  variations from a low of $6.6\pm0.7$~keV at the merging low-entropy,
  high-metallicity, cool core to a high of $22 \pm 6$~keV.  We also
  see a wake in the X-ray surface brightness and deprojected gas
  density caused by the passage of one cluster through the other.
  Archival radio data at 843~MHz reveal diffuse radio emission that,
  if associated with the cluster, indicates the presence of an intense
  double radio relic, hosted by the highest redshift cluster yet.
  \J0102 is possibly a high-redshift analog of the famous Bullet
  Cluster.  Such a massive cluster at this redshift is rare, although
  consistent with the standard \lcdm\ cosmology in the lower part of
  its allowed mass range.  Massive, high-redshift mergers like
  \J0102\ are unlikely to be reproduced in the current generation of
  numerical N-body cosmological simulations.

\end{abstract}


\section{Introduction}

There are currently only a few examples of merging cluster systems in
which there are spatial offsets (on the order of 200-300 kpc) between
the peaks of the total and baryonic matter distributions. Some of
these, like 1E0657$-$56 \citep[the original ``bullet'' cluster at
  $z=0.296$,][]{Markevitch2002}, Abell 2146 \citep[at
  $z=0.234$,][]{Russell2010}, and possibly Abell 2744 \citep[at
  $z=0.308$,][]{Merten2011}, contain in addition a cold, dense
``bullet'' of low entropy gas that is clearly associated with the
merger event. The offsets are due to the differing physical processes
that act on the gas, galaxies, and dark matter.  The gas behaves as a
fluid, experiencing shocks, viscosity, and ram pressure, while the
galaxies and dark matter (or so we posit) are collisionless.  These
bullet systems have been used to offer direct evidence for the
existence of dark matter \citep{Clowe2004,Clowe2006} and to set
constraints on the self-interaction cross-section of the dark matter
particle \citep{Markevitch2004, Randall2008, Bradac2008}.
Additionally, the large merger velocity of 1E0657$-$56 of around
3000~\kms\ \citep{Mastropietro2008,Markevitch2006}, required to
explain the morphology and temperature of the gas from
\chandra\ observations, is much higher than expected in
cosmological simulations.
These merging clusters provide interesting laboratories for several
important topics in astrophysics and cosmology.

Merging cluster systems can display a wide range of mass ratios,
impact parameters, types of merging systems, and times since closest
approach, and clusters evolve with cosmic time. So it is worthwhile to
find new bullet cluster systems, especially at high redshift, to
exploit their different initial and final conditions and learn more
about dark matter and the assembly of clusters through mergers.
Massive clusters ($\sim$10$^{15}\, M_\odot$) become increasingly rare
at high $z$ while their infall speeds are typically higher in the
early Universe \citep[see][]{Lee-Komatsu2010}.

In this paper we report on \J0102, a recently-discovered system
\citep{Marriage2011} which appears to be an excellent new example of a
bullet cluster at redshift $z=0.87$.  It was found by the Atacama
Cosmology Telescope \citep[ACT,][]{Fowler2007} through its strong
Sunyaev-Zel'dovich signal \citep[SZ,][]{SZ1972} , and was confirmed
through optical and X-ray data by \cite{Menanteau-SZ}.  We demonstrate
here that \J0102\ is a rare and exceptional system at a time in cosmic
history when the Universe was only half its current age. Additionally,
it is a contender for being the most massive and X-ray luminous galaxy
cluster at $z>0.6$, likely competitors being CL~J1226+3332
\citep{Maughan2004} at $z=0.89$ with mass\footnote{Throughout this paper 
we distinguish between masses quoted with respect to the average or critical 
density using the ``a'' or ``c'' subscripts.} 
of
$M_{200c}=(1.38\pm0.20)\times10^{15}h_{70}^{-1}M_\odot$
\citep{Jee2009} and \SPT2106\ at $z=1.14$ with
$M_{200a}=(1.27\pm0.21)\times10^{15}h_{70}^{-1}M_\odot$
\citep{Foley2011}.
Therefore, to reflect its exceptional mass and to
recognize the Chilean contribution to ACT we dub the cluster ``El
Gordo,'' which means The Fat/Big One in Spanish.

Throughout this paper we quote cluster masses as $M_{200a}$ (or
$M_{500c}$) which corresponds to the mass enclosed within a radius
where the overdensity is 200 (500) times the average (critical) matter
density, and we assume a standard flat \lcdm\ cosmology with
$\Omega_m= 0.27$ and $\Omega_\Lambda=0.73$, and give relevant
quantities in terms of the Hubble parameter $H_0 = 70
h_{70}$~km~s$^{-1}$ Mpc$^{-1}$. For this cosmology the angular
diameter distance to the cluster is $1617\,h_{70}^{-1}$~Mpc and
1$^\prime$ corresponds to a physical scale of $470\,h_{70}^{-1}$~kpc
at the cluster redshift. Magnitudes are reported in the AB system and
uncertainties are reported for 68\% confidence intervals.

\begin{figure*}[t!]
\centerline{
\includegraphics[width=3.525in]{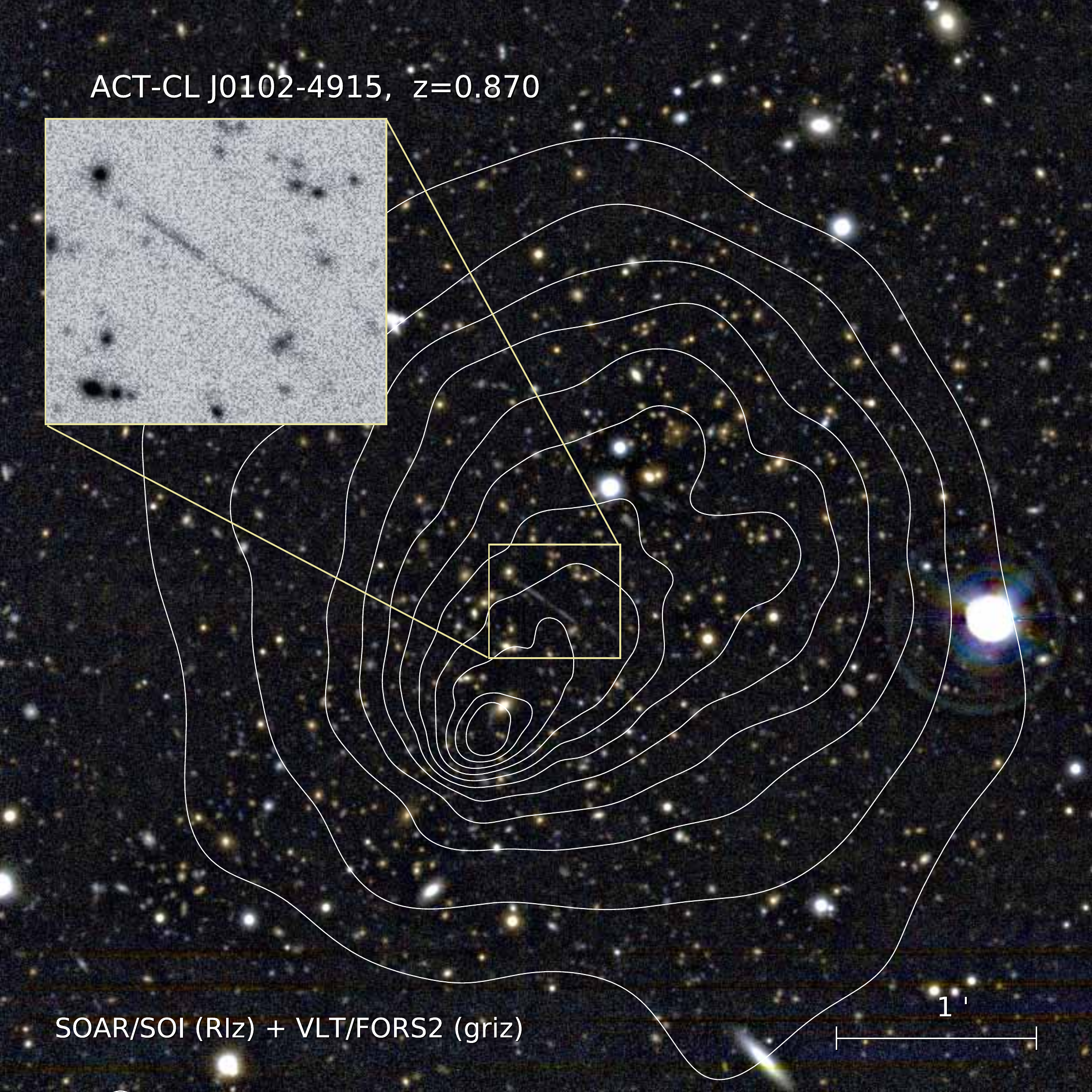}
\includegraphics[width=3.525in]{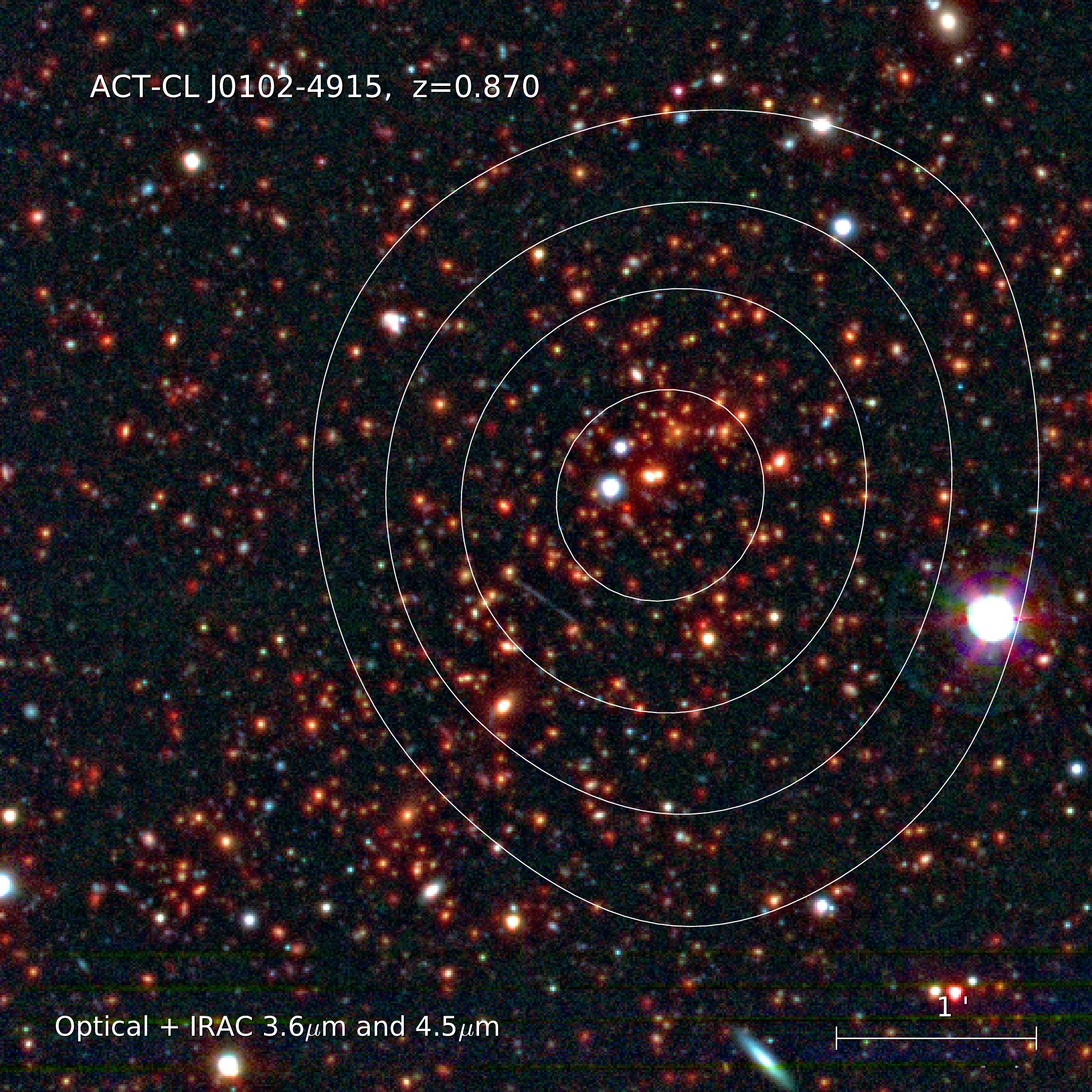}
}
\centerline{
\includegraphics[width=3.525in]{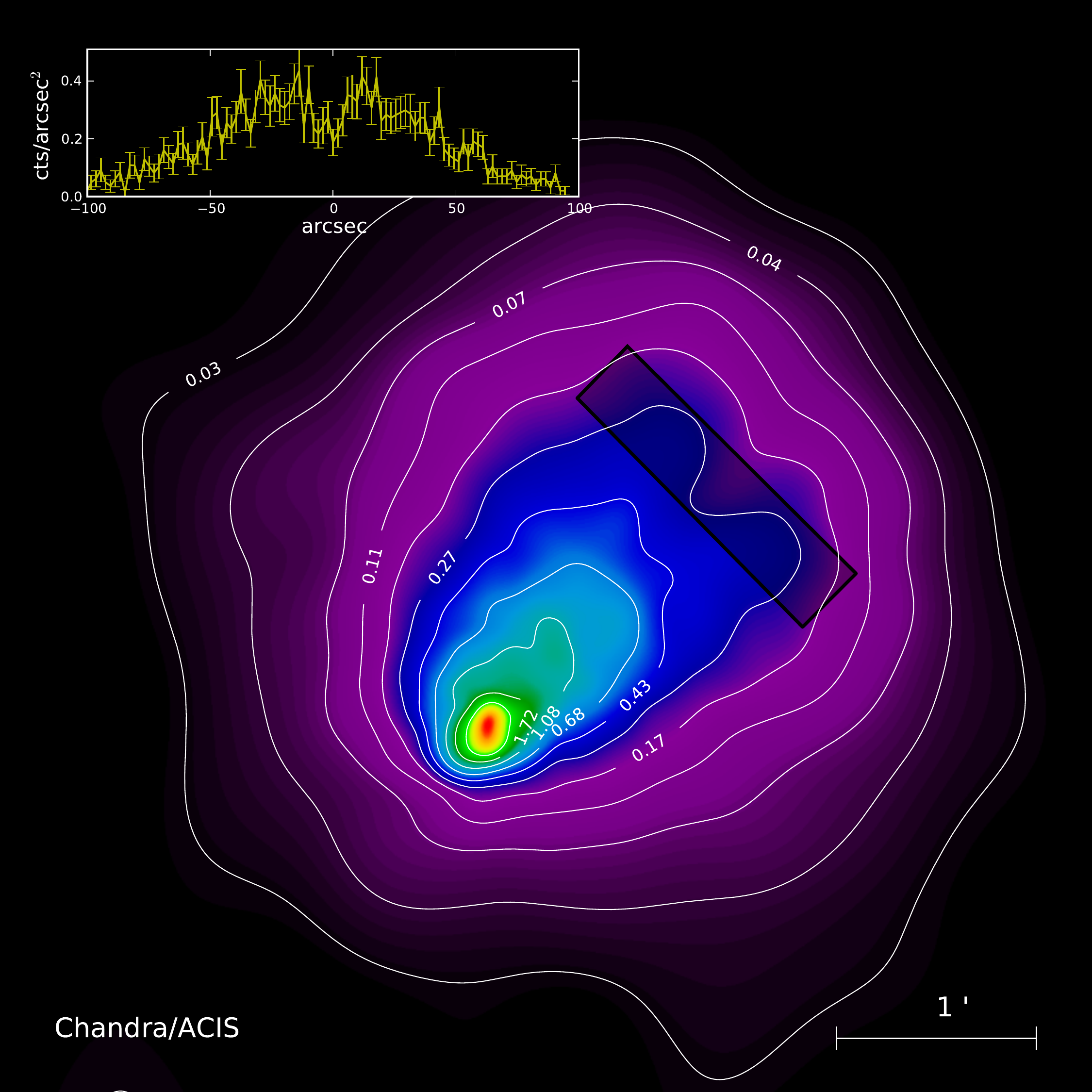}
\includegraphics[width=3.525in]{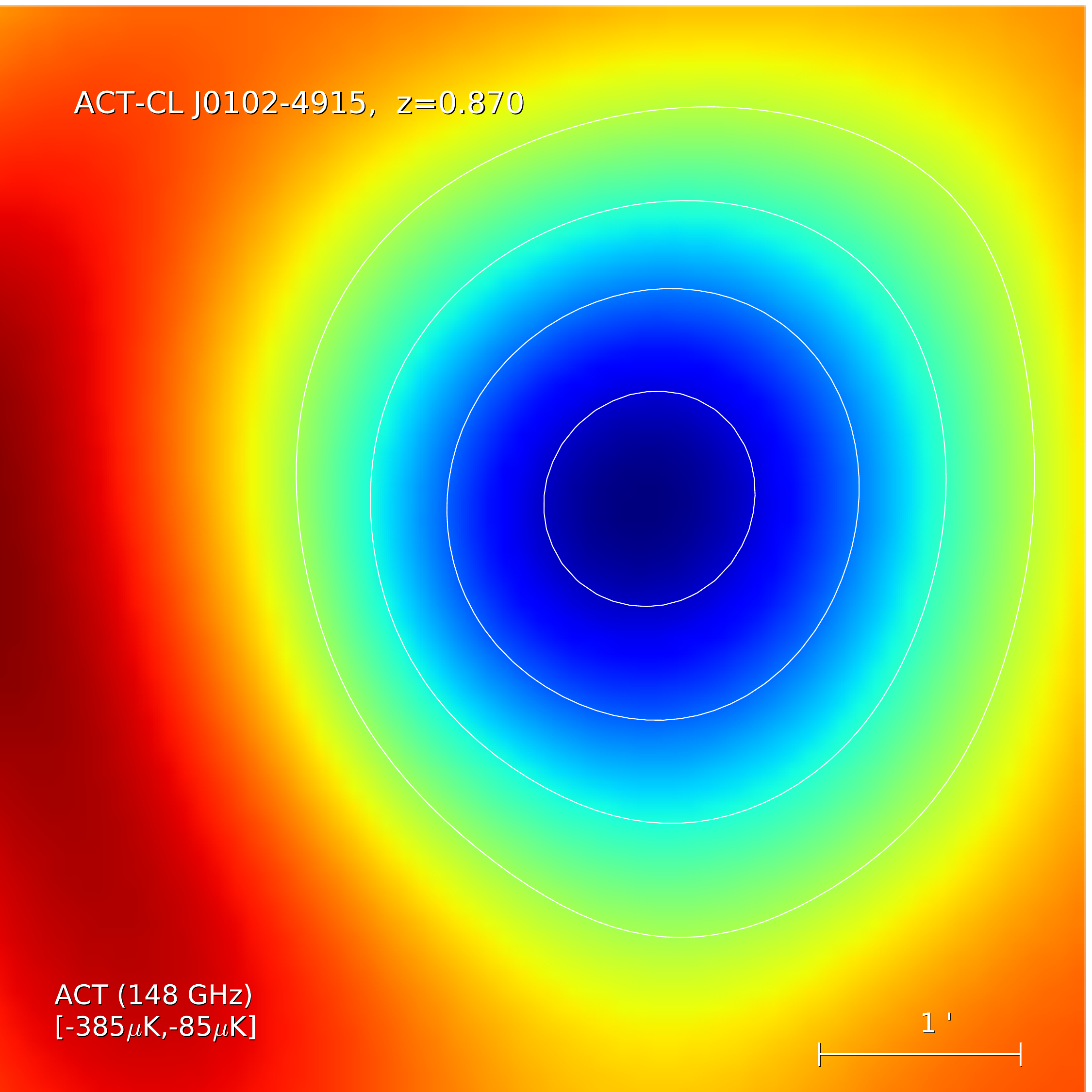}
}
\caption{ The multi-wavelength dataset for \J0102 with all panels
  showing the same sky region. {\em (Upper left)} The composite
  optical color image from the combined $griz$ (SOAR/SOI) and $Riz$
  (VLT/FORS2) imaging with the overplotted \chandra\ X-ray surface
  brightness contours shown in white. The black and white inset image
  shows a remarkably strong lensing arc. {\em (Upper right)} The
  composite color image from the combination of the optical imaging
  from VLT and SOAR and IR from the \spitzer/IRAC \3p6 and \4p5
  imaging. The overplotted linearly-spaced contours in white
  correspond to the matched-filtered ACT 148~GHz intensity maps. {\em
    Bottom left} False color image of the \chandra\ X-ray emission
  with the same set of eleven log-spaced contours between 2.71
  cts/arcsec$^2$ and 0.03 cts/arcsec$^2$ as in the panel above. The
  insert here shows the X-ray surface brightness in a cut across the
  ``wake'' region from the box region shown. {\em Bottom right} ACT
  148~GHz intensity map with angular resolution of $1.\!\!'4$ and
  match-filtered with a nominal galaxy cluster profile, in units of
  effective temperature difference from the mean. The color scale
  ranges from -85 $\mu$K at the edges to -385 $\mu$K at the center of
  the SZ minimum.  In all panels the horizontal bar shows the scale of
  the image, where north is up and east is left.}
\label{fig:cplates}
\end{figure*}

\section{Observations}

We have embarked on an ambitious multi-wavelength follow-up program
aimed at characterizing the baryonic (i.e., gas, stellar) and dark
matter mass components and the SZ-mass scaling relations of the most
massive clusters discovered by ACT via the SZ effect.
As part of this effort we have obtained total mass estimates based on
galaxy dynamics \citep[via measuring spectroscopic redshifts with
8-m class telescopes,][]{Sifon2012}, cluster X-ray emission with
\chandra\ and stellar masses from infrared IRAC/\spitzer\ imaging. In
the following sections we describe the different datasets and preview
the results when applied to this study of \J0102. Images of the cluster
from these various data sources are shown in Fig.~\ref{fig:cplates}.

\subsection{SZ observations}

ACT is a six-meter off-axis telescope designed for arcminute-scale
millimeter-wave observations and it is situated at an elevation of
5190~m on Cerro Toco in the Atacama desert in northern Chile
\citep[see][for a complete description]{Fowler2010,Swetz2011}.
One of the goals of this project is to measure the evolution of
structure in the Universe using massive galaxy clusters detected via
the SZ effect \citep[e.g.,][]{Sehgal2011}.

ACT observations began in late 2007 and to date it has surveyed
two sky areas: one near declination $-55$ degrees (the southern strip)
and the other on the celestial equator.
The 2008 southern observations covered a region of 455~deg$^2$
and recently ACT has analyzed an additional 300 deg$^2$
along the equator \citep[centered on the SDSS Stripe82,][]{DR7}.

\J0102\ was reported by ACT as a particularly strong SZ detection at a
frequency of 148~GHz among 23 high-significance clusters from the 2008
southern survey \citep{Marriage2011}.  In Figure~\ref{fig:cplates}
(lower right panel) we show the filtered ACT 148~GHz intensity maps
for \J0102\ using a matched filtered \citep[e.g.,][]{Haehnelt1996,
  Melin2006}. 

The South Pole Telescope \citep[SPT,][]{SPTref} has reported its most
significant cluster detections on a 2,500 deg$^2$ survey
\citep{Williamson2011}, which overlaps with the ACT 2008 southern
survey. This sample also contains \J0102; it is SPT's most significant
SZ cluster detection to date by nearly a factor of two, with a
comparable beam-averaged SZ decrement to the Bullet cluster
1E0657$-$56.  Among all clusters in the combined ACT and SPT sample,
\J0102\ is exceptional in having the highest SZ signal
\citep[ACT,][]{Marriage2011} or the highest significance SZ detection
\citep[SPT,][]{Williamson2011}.

\subsection{Optical Imaging}
\label{sec:optical}

The initial optical observations of \J0102 were carried out during
December 9--12, 2009 using the SOI camera on the 4.1-m SOAR Telescope
in Cerro Pach\'on (09B-0355, PI: Menanteau) using the SDSS $griz$
filter set with exposure times of 540~s ($6\times90$~s), 720~s
($6\times120$~s), 2200~s ($8\times275$~s) and 2200~s ($8\times275$s)
in $g$, $r$, $i$ and $z$ respectively. These exposure times were twice
the nominal observations for the program as \J0102 was one of the
targeted detections suspected to be a higher-redshift cluster. The run
conditions were optimal and all four nights were photometric with
seeing in the range $0.\!\!\arcsec5$ to $0.\!\!\arcsec9$. A complete
description of the observations and data analysis is provided in
\cite{Menanteau-SZ}.

As part of our 2010B spectroscopic follow-up campaign of the southern
cluster sample, we secured additional deep and wider $Riz$ imaging
(intended for slit mask design) and extensive multi-object
spectroscopy (MOS) with FORS2 on the Very Large Telescope (VLT,
086.A-0425 PI:Infante). FORS2 is a multi-mode (imaging, polarimetry,
long slit and multi-object spectroscopy) optical instrument mounted on
the UT1 (Antu) telescope. We devoted a total of 12~hours of observing
time (service mode) on VLT to \J0102 of which 10~hours were devoted to
MOS and 2~hours to imaging.

In our original confirmation of \J0102 we reported the larger
concentration of galaxies to the NW as the optical cluster counterpart
to the ACT source and used the brightest galaxy it contained as the
cluster center (see Figure~\ref{fig:cplates} top panels). However,
even in the smaller $\approx4.'5\times4.'5$ coverage of the SOAR
images \citep [see Figure~9 from][]{Menanteau-SZ}, we noted a number
of red galaxies with similar photometric redshifts trailing towards
the SE. The lack of coverage and large photometric redshift
uncertainties made it difficult to confidently identify these galaxies
as part of the cluster.

The FORS2/VLT imaging aimed to obtain wider and deeper observations to
confirm the substructure hint in the SOAR data and to secure galaxy
positions for the spectroscopic observations. Our observations were
carried out with the standard resolution of $0.\!\!\arcsec25$/pixel
providing a field size of $6.\!\arcmin8\times6.\!\arcmin8$ with
exposure times of 640~s ($8\times80$~s), 2200~s ($10\times220$~s) and
2300~s ($20\times115$~s) in $R$, $I$ and $z$, respectively. The
optical data were processed using a slightly modified version of the
Python-based Rutgers Southern Cosmology image pipeline
\citep[][]{SCSI,SCSII} that takes us from raw data to galaxy
catalogs. Here we give a brief overview of the steps involved and
refer the reader to the above papers for a complete description. The
initial standard CCD processing steps (i.e. flat fielding, bias
subtraction) were performed via system calls to the ESO Recipe
Execution tool
(EsoRex)\footnote{\url{http://www.eso.org/sci/software/cpl/esorex.html}}.
In order to increase the depth and coverage for \J0102 we co-added the
imaging in the $r$, $i$ and $z$ and $R$, $I$ and $z$ from SOAR and
FORS2 respectively into an {\em \"uber} dataset.
As the FORS2 $RIz$ images were taken under clear but non-photometric
conditions we utilize the SOAR photometric data to calibrate the
photometry of the combined data. All science frames were aligned and
median combined into a common pixel reference frame using SWarp
\citep{SWarp} to a plate scale of $0.\!\!\arcsec2525$/pixel. Source
detection and photometry for the science catalogs were performed using
SExtractor \citep{SEx} in dual-image mode in which sources were
identified on the {\em \"uber} $i-$band images using a $1.5\sigma$
detection threshold, while magnitudes were extracted at matching
locations from all other bands.

\subsection{Optical Spectroscopy and Redshifts}
\label{sec:VLT}

\begin{figure}
\centerline{\includegraphics[width=3.9in]{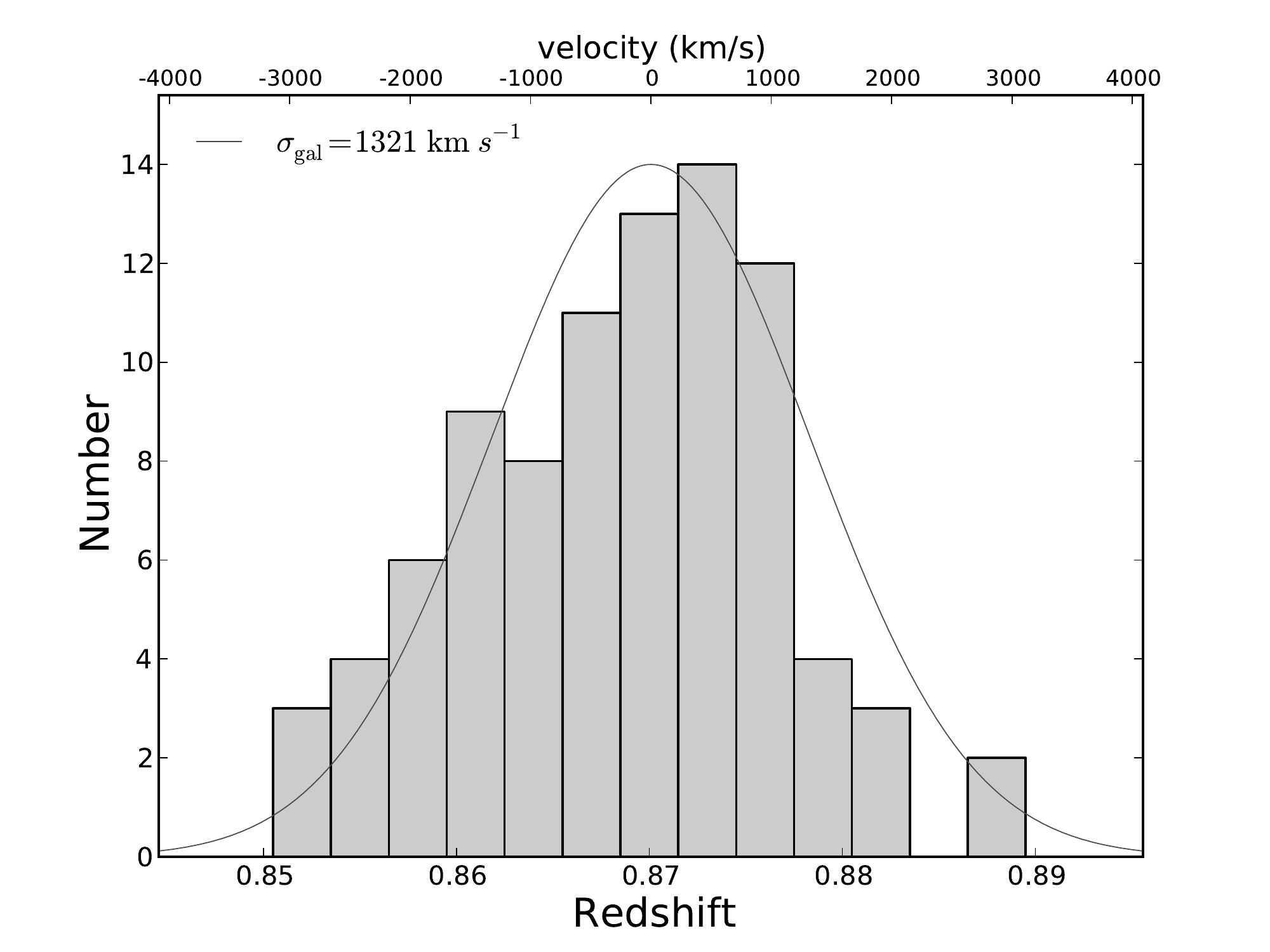}}
\caption{Histogram showing the redshift distribution for the 89 
  cluster member galaxies and the Gaussian with the same
  mean and width $\sigma_{\rm gal}$.}
\label{fig:zhist}
\end{figure}

Based on the possibility that \J0102 was exceptional, we devoted the
bulk (10 of 15 hrs) of the awarded time on VLT in semester 2010B to
securing spectra for galaxies with colors consistent with being
early-type cluster members.
We refer the reader to a companion paper \citep{Sifon2012} where we
fully describe the spectroscopic observing campaign of the 19 clusters
at $z>0.3$ in the ACT southern sample (of which \J0102 was part),
including the data reduction, analysis and redshift
determination. Here we summarize the most salient aspects of this
work.

The observations were executed in January 2011 under photometric
conditions with seeing of $\lesssim0.\!\!\arcsec8$.  The observations
cover the wavelength range $\sim4000-8000$\AA\ using the GRIS
300+11~Grism at 1.62~\AA/pixel and resolution
$\lambda/\Delta\lambda~\sim660$ at the central wavelength.
At the cluster redshift of $z=0.87$, the [O~II]3727\AA\ emission
lines, the CaII (K-H, $3950$\AA) (which is the spectral signature of
elliptical galaxies), plus other absorption lines such as the G-band
($4300$\AA) as well as the 4000\AA\ break were located in the spectral
range of the grism. These features facilitated the determination of
galaxy redshifts, galaxy cluster membership, and the galaxy velocity
dispersion of the cluster.
Integration times were 40 minutes per mask in order to maximize the
number of galaxies while obtaining the necessary signal-to-noise ratio
in the relevant lines. We observed a total of five FORS2/MXU masks in
this configuration, providing redshifts for a total of 123 objects
of which 89 were cluster members.

The redshifts were measured by cross-correlating the spectra with
galaxy spectral templates of the Sloan Digital Sky Survey (SDSS) Data
Release 7 \citep{DR7} using the RVSAO/XCSAO package for IRAF
\citep{RVSAO}.
Most of these member galaxies are ellipticals with no emission lines,
and only a few emission line galaxies have been found to belong to the
cluster.

Cluster membership was determined using a rest-frame cut in velocity
space of $4000\,\mathrm{km\,s^{-1}}$, and then applying the shifting
gapper method of \cite{Katgert-96} in order to remove galaxies that
lie outside a given gap in velocity space from the main body of
galaxies. In particular, we defined the main body in bins of 15
galaxies and used a velocity gap of $500\,\mathrm{km\,s^{-1}}$.
We used the well-accepted biweight estimator \citep{Beers-90} to
estimate both the mean redshift of the sample, $z=0.87008\pm0.00010$,
and the velocity dispersion\footnote{As customary the quoted uncertainties for $z$ and
  $\sigma_{\rm gal}$ only reflect statistical errors from 89 objects
  and do not include systematic effects.}  
of the cluster, $\sigma_{\rm gal}=1321 \pm 106$~km~s$^{-1}$.
Errors were estimated via bootstrap resampling.
The velocity dispersion of \J0102 is larger than all other
clusters in the ACT sample \citep{Menanteau-SZ,Sifon2012} and also
larger than SPT-CL~2106$-$5844 which is the largest in the full 2,500
deg$^2$ SPT sample \citep{Foley2011,Williamson2011}. Using only the
passive galaxies (i.e., objects with no emission lines) in \J0102, we
obtained a mean redshift of $z=0.86963\pm0.00012$ with a velocity
dispersion of $1280\pm108$~km~s$^{-1}$.

In Figure~\ref{fig:zhist} we show the redshift distribution
for all galaxy members from the FORS2 spectra. We see no
indication of substructure along the line of sight, confirming
the conclusion of  \cite{Sifon2012} which used  the \cite{Dressler1988}
test. However, as we will discuss in Sec.~\ref{sec:structure} 
we find substantial evidence for spatial
substructure.

\subsection{Chandra X-ray Observations}
\label{sec:xco}

\begin{figure}
\centerline{\includegraphics[width=3.2in]{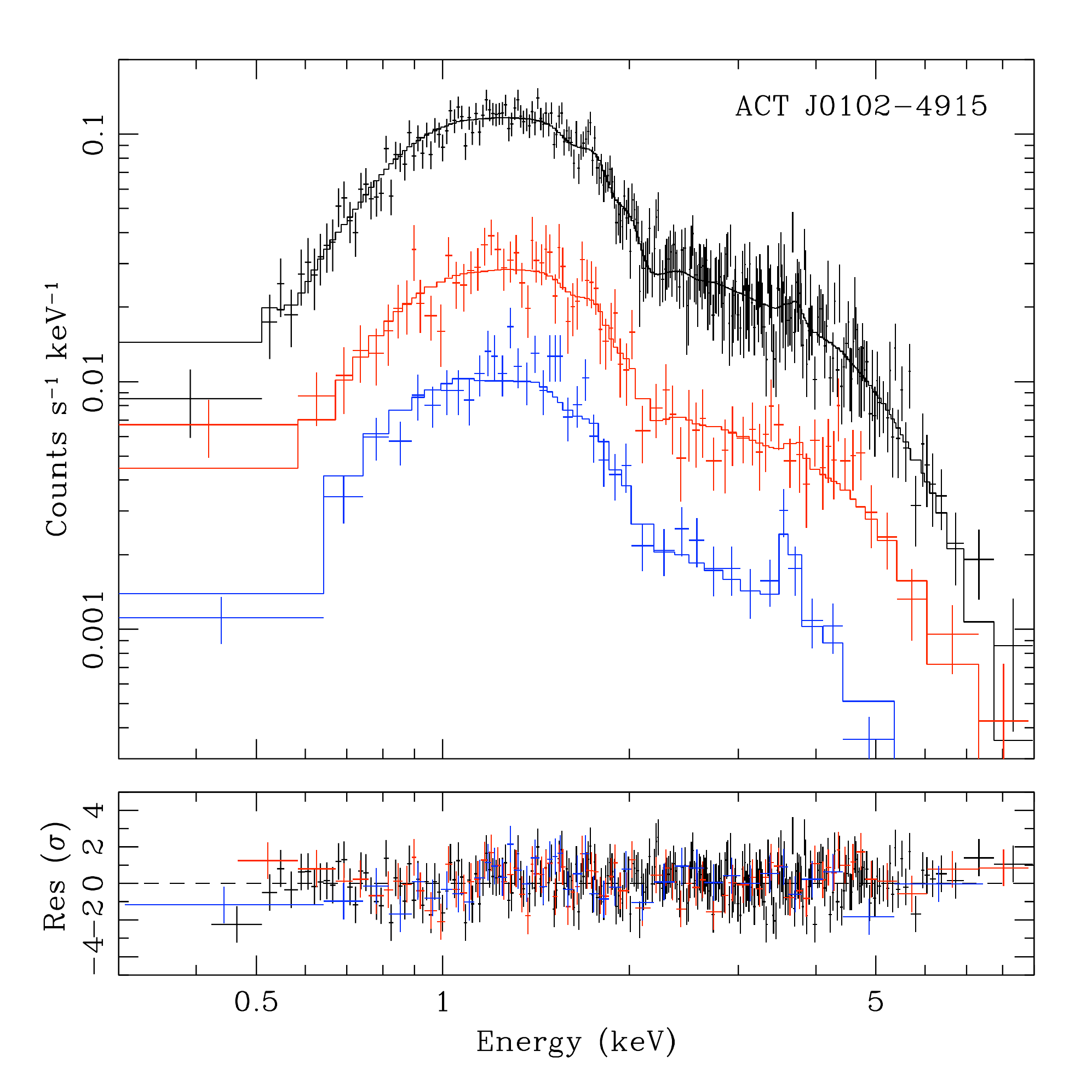}}
\caption{\chandra\ X-ray spectra of \J0102\ for the total cluster in
  black, the cool bright peak (region 1 in Fig.~\ref{fig:regions}) in
  blue, and the highest temperature region (region 5) in red.  The
  bottom panel shows the residuals between the best fit models
  (histograms in top panel) and data. 
}
\label{fig:xco}
\end{figure}

\begin{figure}
\centerline{
\includegraphics[width=3.6in]{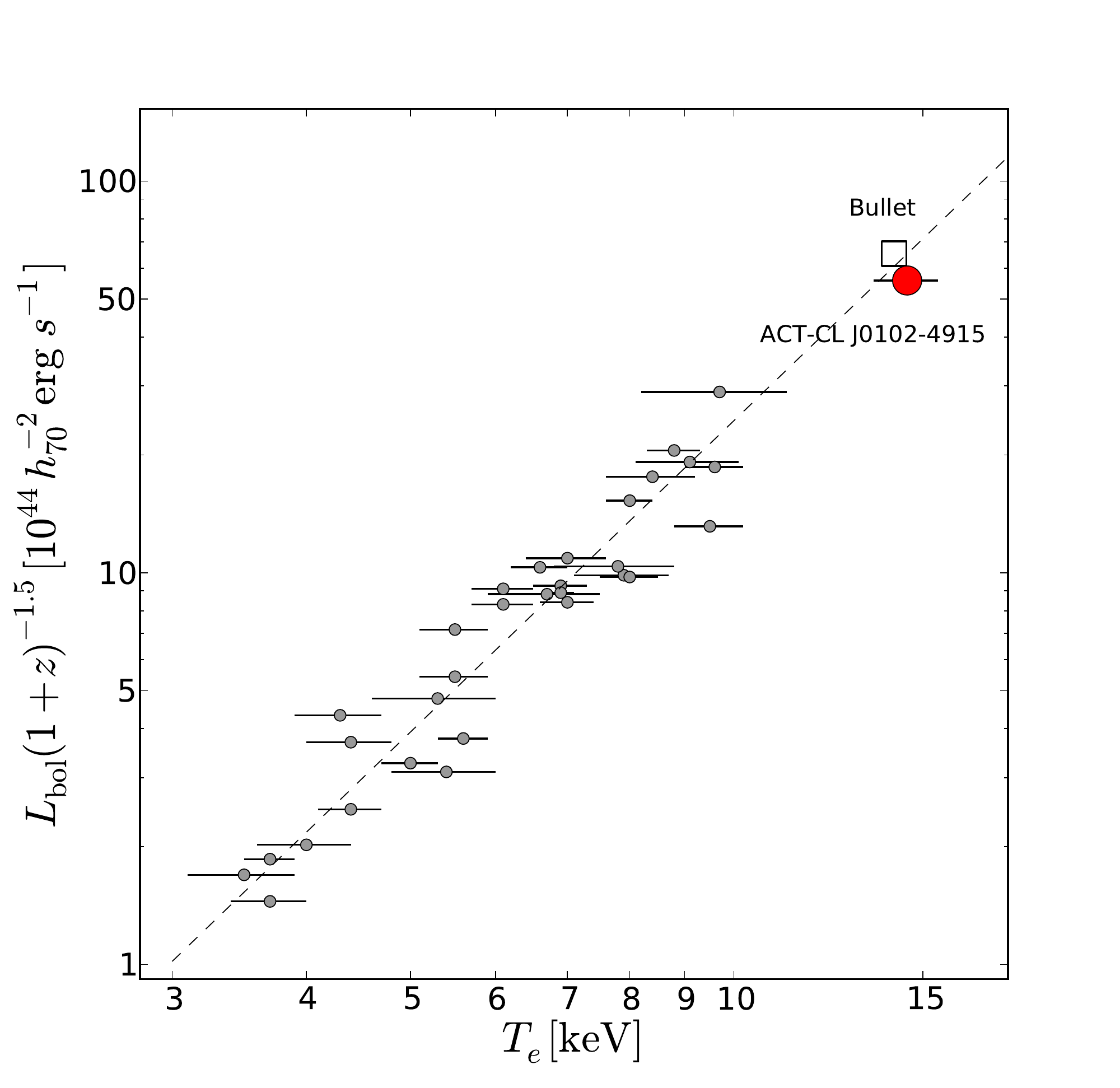}
}
\caption{The X-ray bolometric luminosity vs.\ temperature for a sample
  of well-studied clusters taken from \cite{Markevitch1998}. The
  Bullet cluster (1E0657$-$56) is the open square point at high
  temperature and luminosity from \cite{Markevitch2006}. \J0102\ is
  the red circle in the same region of the plot.}
\label{fig:L-T}
\end{figure}

We were awarded 60 ks of \chandra\ observing time on this target
during cycle 12 (PI: Hughes, ObsId: 12258) as part of our follow-up
effort on the SZ clusters found in the ACT 2008 455-deg$^2$ survey
\citep{Menanteau-SZ}. The \chandra\ observation was carried out on
January 26--27, 2011 for an effective exposure of 59,260~s using the
ACIS-I array. Figure~\ref{fig:cplates} (bottom left panel) shows the
surface brightness of the cluster after point source removal, exposure
correction, and adaptive kernel smoothing.

The cometary appearance of the X-rays is remarkable, extending even to
the apparent presence of two ``tails'' extending off toward the NW.
Orthogonal slices of the X-ray surface brightness show that these
``tails'' arise from significant depressions 
in the X-ray intensity, at a level of 20--40\%, 
in a band extending from about 1$^\prime$
NW of the compact peak emission region and continuing off to the NW
out to the faint parts of the cluster.  The depression is about
35$^{\prime\prime}$ ($270\,h_{70}^{-1}$ kpc) wide. We refer to
this feature as the ``wake''.

The integrated \chandra\ spectrum is shown in Figure~\ref{fig:xco}
(top black points). The extraction region for this spectrum excludes
the central peak emission (from within the innermost contour in
Fig.~\ref{fig:cplates}) and extends out to the outermost contour in
Fig.~\ref{fig:cplates}, which is at a radius of around $2.\!'2$
($1.0\,h_{70}^{-1}$ Mpc). An absorbed {\tt phabs*mekal}
model yielded a best-fit (source frame) temperature of $kT=14.5\pm1.0$
keV, metal abundance $0.19\pm 0.09$ with respect to solar values, and
a bolometric luminosity of $L_{\rm bol}=13.6\times10^{45}\,
h_{70}^{-2}$~erg~s$^{-1}$.  Figure~\ref{fig:L-T} shows the $L_{\rm
  bol}$-$T_X$ relation with \J0102\ added as the red point \citep[note
  that a redshift correction factor of $(1+z)^{-1.5}$ was applied to
  $L_{\rm bol}$ following][]{Vikhlinin2002}. The nearby point on this
figure is the closest comparison cluster, 1E0657$-$56.  Similarly, the
X-ray luminosity of \J0102 in the 0.5--2.0 keV band is
$L_X=2.19\pm0.11\times10^{45}\,h_{70}^{-2}$erg~s$^{-1}$.

\begin{figure}
\centerline{\includegraphics[width=4.2in]{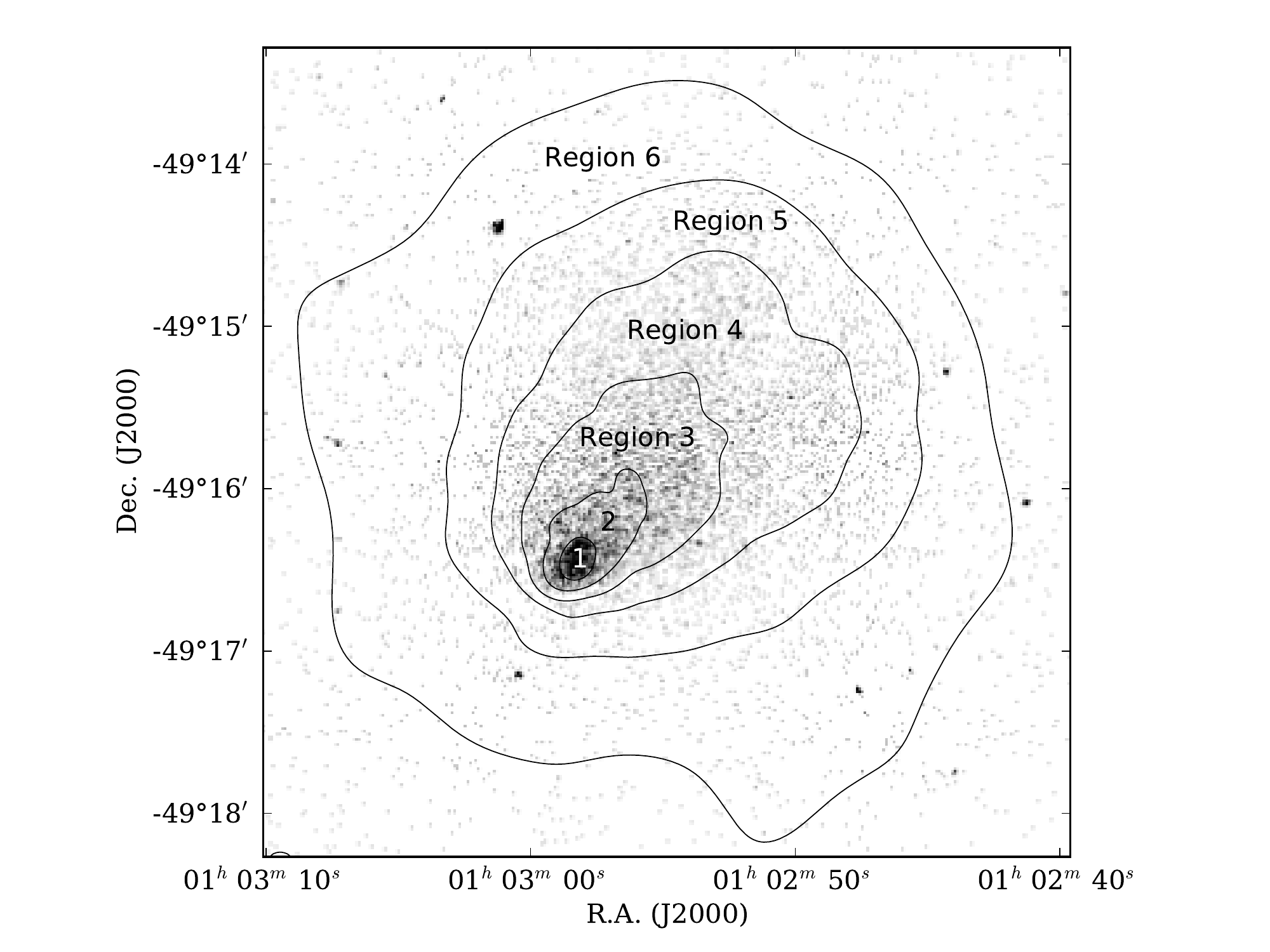}}
\includegraphics[width=3.2in]{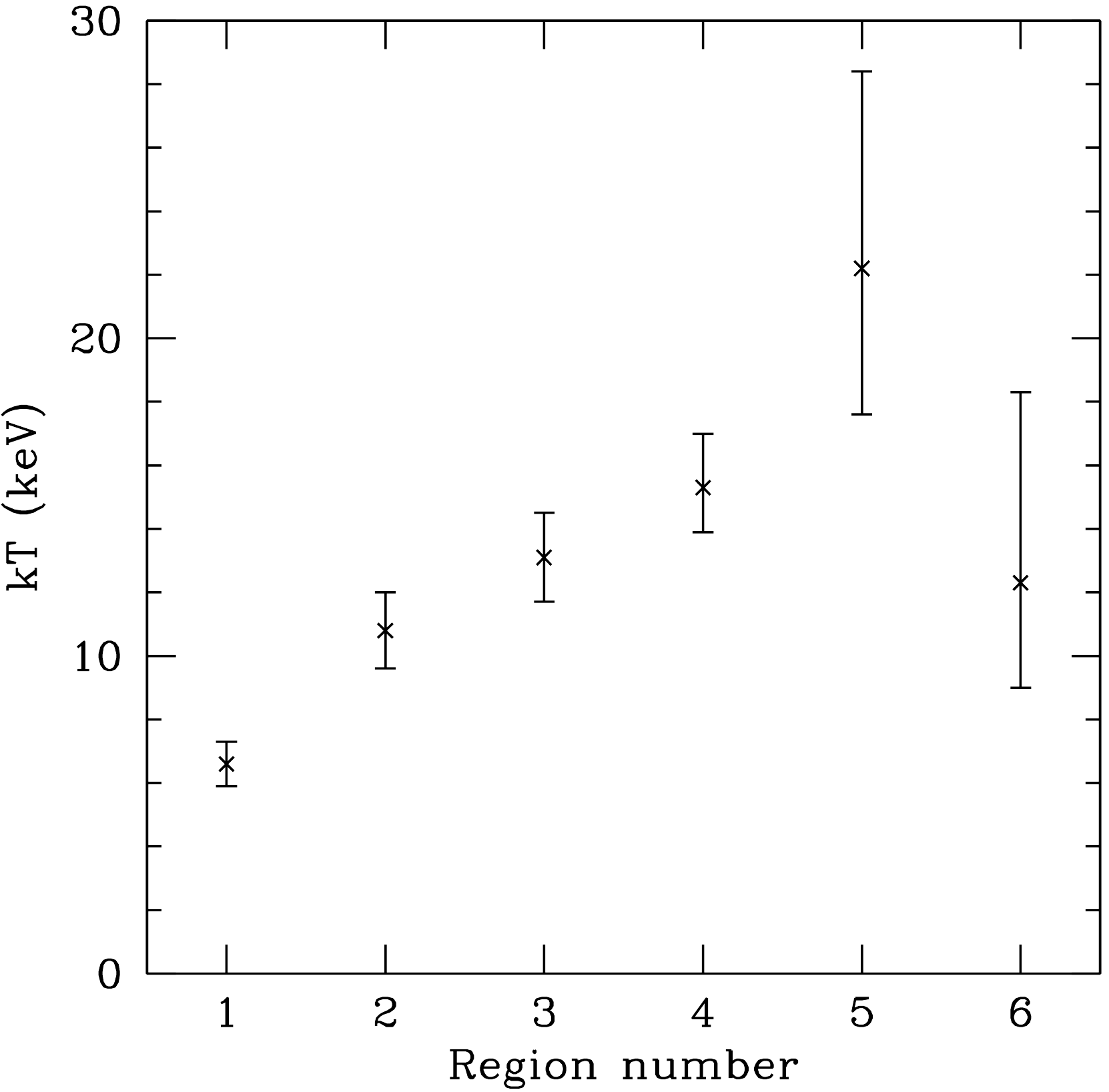}
\caption{ ({\it Upper panel}) \chandra\ X-ray image showing raw
  detected counts and the spectral extraction regions. ({\it Lower
    panel}) Best-fitted X-ray temperatures plotted versus spectral
  extraction region (see key in figure to left). Error bars are
  1-$\sigma$.  Region 1, corresponding to the bright peak,
  is the coolest region.}
\label{fig:regions}
\end{figure}

\begin{figure*}
\centerline{
\includegraphics[width=3.3in]{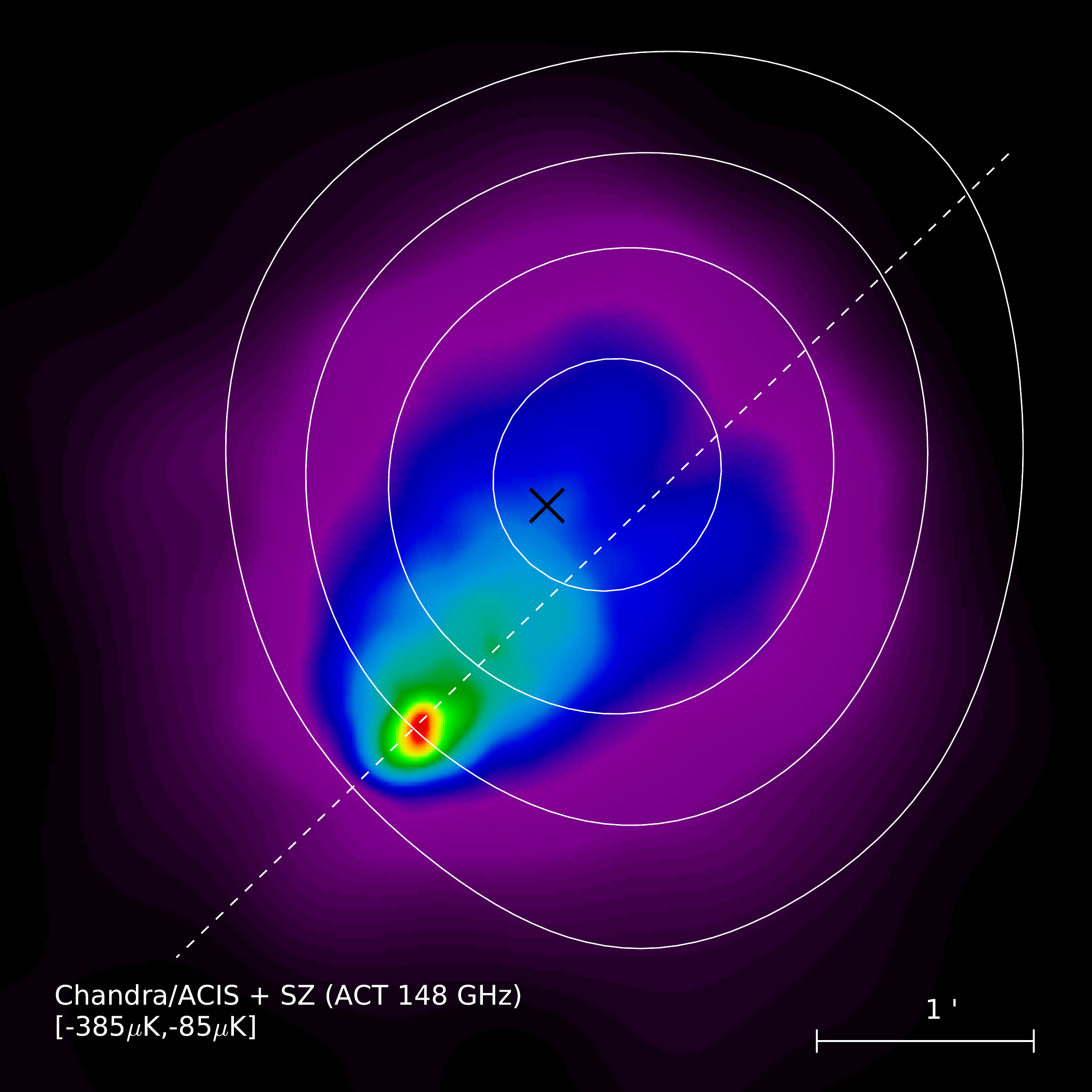}
\includegraphics[width=3.3in]{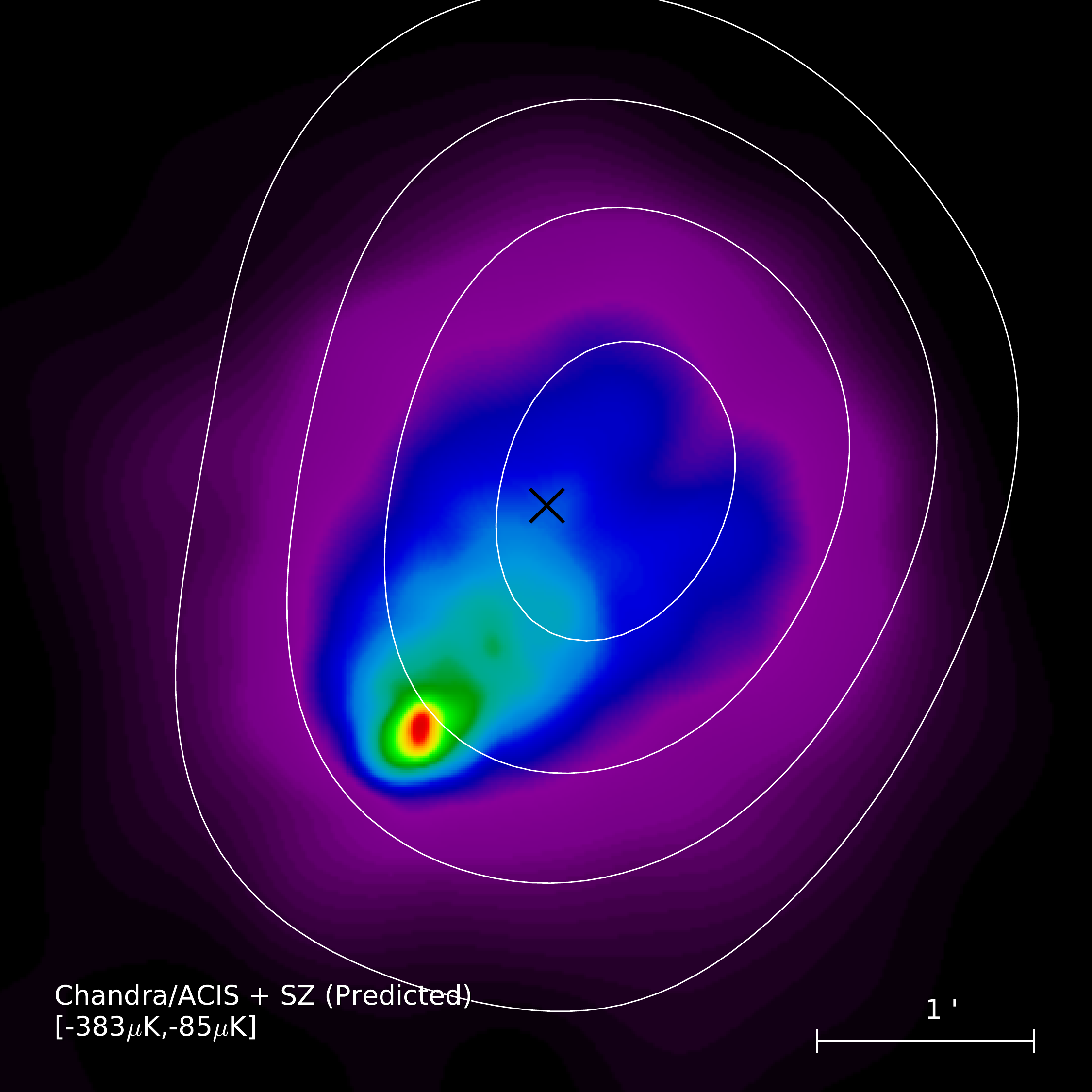}
}
\caption{The X-ray \chandra\ emission overlaid with the ACT 148~GHz
  data contours from Fig.~\ref{fig:cplates} {\em (left)} and the
  predicted SZ signal at 148~GHz from the deprojected X-ray data
  {\em(right)}. The black cross represents the SZ position reported by
  \cite{Williamson2011} for \J0102. The contour levels are linearly
  spaced between the limits shown in brackets on each panel. The
  minimum central value of the ACT and predicted signals agree to
  within $\sim$15\%. The white dotted line shows the symmetry axis
  chosen for the deconvolution.}
\label{fig:sz-cxo}
\end{figure*}

The modest count rate of \J0102\ (i.e., 0.2 s$^{-1}$ corresponding to
a total number of detected X-ray photons of nearly 12,000) limits the
number of regions into which the cluster can be subdivided for
spatially resolved spectroscopy. We perform this measurement by
dividing the total emission into roughly equal parts as shown in
Figure~\ref{fig:regions}. Spectra were extracted and fitted to the
same absorbed thermal model as before; the best-fit temperatures are
plotted vs.\ region number in Fig.~\ref{fig:regions}. The coolest
(region 1, $kT=6.6\pm0.7$ keV) and the hottest (region 5,
$kT=22^{+6}_{-5}$ keV) spectra are plotted in Figure~\ref{fig:xco} in
blue and red, respectively.  Note the strong Fe K line (observed
at $E=3.56$~keV) in the spectrum of the coolest region, which yielded a
rather high metal abundance of $0.57\pm 0.20$.  Elsewhere the Fe K
line is not detected strongly, pointing to abundance variations in the
cluster gas.

We see that the peak emission region in \J0102\ is an X-ray--bright,
relatively cool, ``bullet'' of low entropy gas like in 1E0657$-$56.
The enhanced Fe abundance suggests it was the core of a cool-core
cluster that has, so far, survived a high velocity merger with another
system. The cluster's high overall gas temperature and spatial
temperature variations (Fig.~\ref{fig:xco}) argue for significant
amounts of shock heating of the gas.  The steep fall-off in the X-ray
surface brightness toward the SE, as well as the ``wake'' in the main
cluster gas toward the NW, clearly indicate that the bullet is moving
toward the SE.

The two lower panels of Fig.~\ref{fig:cplates} show that the SZ and
X-ray peaks are displaced by about $1.\!\arcmin2$.  Similar offsets
between the peak SZ and X-ray emission have also been reported in the
bullet-like cluster Abell 2146 at $z=0.23$, for which \chandra\ and
Arcminute Microkelvin Imager (AMI) observations at 13.9--18.2~GHz show
a remarkably similar separation of the SZ and X-ray signal
distributions \citep[see][Figures 13 and 15]{AMI2011}.  Two factors
contribute to this.  First, X-ray emission is more sensitive to denser
parts of the cluster, since X-ray surface brightness is proportional
to $\int n_e^2 T_e^{\alpha}dl$, where $n_e$ and $T_e$ are the electron
number density and temperature and $\alpha\simeq 0.1$ (for
band-limited X-ray emission at high temperature), while the SZ signal
is proportional to the line-of-sight gas pressure $\int n_e T_e
dl$. Second, the combination of \chandra\ and ACT's widely different
angular resolutions ($< 1^{\prime\prime}$ vs.~$1.\!\!'4$) is
particularly relevant for observations of merging clusters with
complex gas density substructure.

In order to investigate these differences we developed a new technique
for deprojecting X-ray images that, unlike traditional methods
utilizing azimuthally-averaged profiles, takes account of the full
observed two-dimensional structure.  The process assumes rotational
symmetry about a user-defined axis and independently deprojects each
column of pixels perpendicular to that axis.  For \J0102\ we chose the
symmetry axis so that it runs through both the peak X-ray emission at
the bullet and the surface brightness depression toward the NW.  The
axis is centered on R.A.=01:02:54.9, Dec.=$-$49:15:52.5 and oriented
at a position angle of 136$^\circ$ (see Figure~\ref{fig:sz-cxo}).
Beyond a certain radius from the center ($1.\!'65$ in the case of
\J0102), where the surface brightness is background dominated, we used
an analytic $\beta$-profile fit to the azimuthally-averaged data to
deproject the cluster emission smoothly to large radii.  Interior to
this radius, using the adaptively-smoothed, 0.5--2.0 keV band image,
we deproject the X-ray surface brightness in strips perpendicular to
the symmetry axis.  Deprojection of the cluster emission is done in
the usual manner (``peeling the onion'') using the best-fit integrated
spectral parameters given above to relate the X-ray emission to the
gas density of each deprojected shell. We deproject the north and
south parts of the cluster separately, but interpolate the density
value across the cluster when ``peeling'' back any individual layer
(to avoid a discontinuity at the symmetry axis).  In general this
process is not guaranteed to produce positive definite values (as
physically required), so we force the computed density to be zero in
cases where the deprojection would result in a negative (or more
precisely, imaginary) value.  We quote electron density values using
an electron to proton number ratio of $n_e/n_p = 1.2$ and for mass
densities we assume a mean mass of $1.4 m_p$.

The deprojection results in an estimate of the three-dimensional
distribution of the gas density in the cluster. From this we compute
the predicted SZ signal at 148~GHz.  Since the SZ depends linearly on
the gas temperature, as a first approximation, we use the fitted X-ray
temperatures in the 6 regions shown in Fig.~\ref{fig:xco} to generate
a temperature-corrected SZ map of the cluster. This was done by
computing the ratio of expected SZ signal \citep{Nozawa1998} for the
observed temperatures in each of the six regions to the average
cluster temperature and multiplying the single temperature SZ map by
the appropriate ratio for the region. The multiplicative ratios vary
from a value of 0.5 (region 1) to 1.6 (region 5). A more refined
calculation would take into account the deprojection of the X-ray
temperatures, however, for our purposes here (i.e., to study the
centroid differences), the simpler approximation is sufficent.

This temperature-corrected SZ map is then convolved with the ACT beam
model \citep{Hincks2010}, added to a source-free region of the ACT map
in the vicinity \J0102\ (to include similar CMB and intrumental
noise), and then is filtered with the same procedure used to detect
\J0102.  In Figure~\ref{fig:sz-cxo} we show contours of the predicted
SZ signal overlaid on the \chandra\ data (right), next to an identical
plot with contours of the ACT data (left). The resulting centroid
difference between the peak of the SZ signal and the simulated SZ map
is $6\arcsec$, which is within the expected positional error
circle. Furthermore the intensity of the predicted SZ signal agrees to
within $\sim$15\% with the ACT data.
To conclude, we note that the offset between the peak X-ray and SZ
signals is predominantly a result of the different sizes of the beams
between \chandra\ and ACT (at least a factor of 60).

\subsection{\spitzer/IRAC Imaging}
\label{sec:irac}

As part of a larger warm-phase \spitzer\ program (PID: 70149,
PI:Menanteau) which aims to characterize the stellar mass content of
clusters in the ACT sample (Hilton et al., in prep.), in August 2010 we
obtained IRAC \3p6 and \4p5 observations of \J0102. Our observations
were designed to provide coverage out to the virial radius for
clusters at $z > 0.4$, using a $2 \times 2$ grid of IRAC pointings
centered on the cluster position. A total of $10 \times 100$~s frames
were obtained in each channel at each grid position, using a large
scale cycling dither pattern. The basic calibrated data (BCD) images
were corrected for pulldown using the software of Ashby \&
Hora \footnote{See
  \url{http://irsa.ipac.caltech.edu/data/SPITZER/docs/dataanalysistools/
    tools/contributed/irac/fixpulldown/}} and then mosaiced using
MOPEX \citep{Makovoz2005} to give maps which are $\approx 13\arcmin$
on a side with $0.\!\!\arcsec6$ pixel scale. The maps for each channel
were then registered to a common pixel coordinate system. By inserting
synthetic point sources, we estimate the 80\% completeness depths of
the final maps to be $\approx 22.6$ mag (AB) in both channels.

Matched aperture photometry was performed on the IRAC maps using
SExtractor in dual image mode, using the \3p6 channel as the detection
band. We measure fluxes through $4\arcsec$ diameter circular
apertures, which are corrected to estimates of total magnitude using
aperture corrections (as measured by \citealt{Barmby2008}) of
$-0.35\pm0.04$, $-0.37\pm0.04$ mag in the \3p6, \4p5 channels
respectively. The photometric uncertainties were scaled upwards by
factors of 2.8, 2.6 in the \3p6, \4p5 channels respectively, in order
to account for noise correlation between pixels introduced in the
production of the mosaics which is not taken into account in the
SExtractor error estimates.  These scaling factors were determined
using the method outlined in \citet{Barmby2008}. Finally, the
uncertainties in the aperture corrections were added to the
photometric errors in quadrature.

\section{Analysis and Results}

\subsection{Mass estimation}

Here we estimate the mass for \J0102\ using all the mass proxies
available from the X-ray data and optical spectroscopy. We present
values in Table~\ref{tab:mass} and discuss them in detail in the next
sections.  In all cases we quote uncertainties that include 
measurement errors as well as the uncertainties in scaling relations
between observables and cluster mass.
relation. For the dynamical and X-ray masses, scaling to the common mass
estimator $M_{200a}$ carries additional uncertainty that we quantify
below and include in the Table.

\subsubsection{Dynamical Mass}
\label{sec:dmass}

Here we report in more detail the dynamical mass estimates for \J0102
calculated by \cite{Sifon2012}.
The velocity dispersion of the cluster is $\sigma_{\rm gal}=1321 \pm
106$~km~s$^{-1}$. This matches the largest reported dispersion of all
known clusters at $z>0.6$ from RX~J0152.7$-$1357 at $z=0.836$ and
$\sigma_{\rm gal}=1322^{+74}_{-28}$~km~s$^{-1}$ \citep{Girardi2005},
although note that the quoted velocity dispersion is for a
well-studied merging system which is resolved into two components of
considerably lower mass.
We use the $M_{200c}$-$\sigma_{\rm DM}$ scaling relation from
\cite{Evrard2008}
  to convert the measured velocity dispersion into a dynamical mass
  estimate,
\begin{equation}\label{eq:mass}
 M_{200c} = \frac{10^{15}}{0.7h_{70}(z)}\left(\frac{\sigma_{\rm DM}}{\sigma_{{\rm DM}, 15}}\right)^{1/{\alpha}}M_\odot,
\end{equation}
where $h_{70}(z)= h_{70}\sqrt{\Omega_m(1+z)^3 + \Omega_{\Lambda}}$,
$\sigma_{{\rm DM},15} = 1082.9\pm4.0$~km~s$^{-1}$, $\alpha = 0.3361\pm0.0026$
and $\sigma_{\rm DM}$ is the velocity dispersion of the dark matter
halo. The latter is related to the observed galaxy velocity
dispersion by the velocity bias parameter, $b_v = \sigma_{\rm
  gal}/\sigma_{\rm DM}$. The latest physically motivated simulations
\citep[see][and references therein]{Evrard2008} indicate that 
galaxies are essentially unbiased
tracers of the dark matter potential, $\langle b_v
\rangle=1.00\pm0.05$.
Using a bias factor of $b_v=1$ for the velocity dispersion for all
galaxies, we obtain a dynamical mass of $M_{200a,{\rm dyn}} =
1.86_{-0.49}^{+0.54}\times 10^{15}\,h_{70}^{-1}\, M_\odot$. Similarly, using only the
passively evolving galaxies without emission lines, we obtain $M_{200a,{\rm dyn}} =
1.72_{-0.47}^{+0.50}\times 10^{15}\,h_{70}^{-1}\, M_\odot$. 
We note that the \cite{Evrard2008} scaling uses $M_{200c}$ as the mass
within a radius where the overdensity is 200 times the {\em critical}
density of the Universe, so the observed mass values reported in
\cite{Sifon2012} have been scaled from critical to average density using
$M_{200a} = 1.16_{-0.03}^{+0.04} \times M_{200c}$. 
This conversion factor was derived using a
\cite{NFW} mass profile (hereafter NFW) and the concentration-mass
relation, $c(M,z)$, from simulations \citep{Duffy2008} at $z=0.87$ for
the mass of the cluster. The reported uncertainties in the conversion
factor reflect the $\sigma_{{\rm log}\,c}=0.15$ scatter in the
log-normal probability distribution of $c(M,z)$.
The radius $r_{200} = 2111 \pm 189\,h_{70}^{-1}$~kpc was also calculated using
$M_{200a}$ and assuming spherical symmetry, using the average density
of the Universe at redshift $z$, $\bar{\rho}(z) =\Omega_m(1+z)^3 3H_0^2/\left(8\pi
  G\right)$.

\subsubsection{X-ray Mass}
\label{sec:xmass}

We estimated the cluster mass from several X-ray mass proxies 
using the scaling relations in \cite{V09} and \cite{KVN}.
For the gas temperature and luminosity we essentially follow the
prescriptions in \cite{V09} to apply those scaling laws to the X-ray
data.  We obtain masses of $M_{500c} = 1.39_{-0.19}^{+0.20}\times
10^{15}\,h_{70}^{-1}\, M_\odot$ and $M_{500c} =
1.40_{-0.36}^{+0.44}\times 10^{15}\,h_{70}^{-1}\,M_\odot$ from the
$M_{500c}$-$T_X$ and $M_{500c}$-$L_X$ relations, respectively.  We
refrain, however, from applying the 17\% "merging cluster mass boost"
as implemented by \citet{V09}, since, given the exceptional nature of
\J0102, it is not obvious that such a correction factor is
appropriate.  Due to the complex substructure of the cluster, for the
$M_{\rm gas}$-based relations ($Y_X$ and $M_{\rm gas}$), we use the 2D
deprojection to generate an azimuthally-average radial gas mass
profile about the cluster center, which agrees well (within $\sim$15\%) with the
azimuthally-averaged core-excised surface brightness profile.  For the
$Y_X$ mass proxy we obtain a mass of $M_{500c} =
1.55_{-0.15}^{+0.15}\times10^{15}\,h_{70}^{-1}\,M_\odot$.

Since all of these mass proxies are potentially sensitive to merger
boosts on $T_X$ and $L_X$, we also explored using $M_{\rm gas}$, which
is intrinsically unaffected by merger boosts.  However, the
X-ray--derived measurement of $M_{\rm gas}$, which comes from the
reconstructed density by deprojecting the surface brightness, is
sensitive to clumping. This can in principle result in a bias,
depending on whether clusters are indeed more highly clumped during
mergers.  In the case of \J0102, the excellent agreement between our
predicted SZ signal from the \chandra\ data and the actual mm-band ACT
data suggests that gas clumping is not significant.  We use the
scaling law for $M_{500c} - M_{\rm gas}$ at redshift $z=0.6$ from
\cite{KVN}, which yields a value of $M_{500c} =
1.67_{-0.20}^{+0.20}\times 10^{15}\,h_{70}^{-1}\,M_\odot$ and implies
a gas mass fraction $f_{\rm gas} = 0.133$. 
The agreement among the four X-ray mass proxies is excellent: all are
consistent within their 1-$\sigma$ ranges with a single mass value.

Our $M_{200a}$ X-ray mass estimates are given in Table~\ref{tab:mass}
using a conversion factor of $1.85_{-0.20}^{+0.30}$ from $M_{500c}$ to
$M_{200a}$. This factor is derived using the same procedure as in
Section~\ref{sec:dmass}.

\subsubsection{SZ Mass}

We estimate the SZ-derived mass of this cluster using the peak Compton signal for
the brightest 0.\!'5 pixel ($yT_{\rm CMB}$) of $490\pm60\mu$K as published in
\cite{Sehgal2011}.
We combined this measurement with a fiducial scaling relation between
$yT_{\rm CMB}$ and $M_{200a}$ derived from simulations and calibrated by X-ray
observations with best-fit parameters given in that same work. Using
this scaling relation, we find the mass of this cluster to be
$M_{200,{\rm SZ}}=1.64_{-0.42}^{+0.62}\times10^{15}\,h_{70}^{-1}
M_{\odot}$.  This mass is consistent with the independently-derived
mass reported by SPT \citep{Williamson2011} of $M_{200,{\rm
    SZ}}=(1.78\pm0.64)\times10^{15}\,h_{70}^{-1}M_\odot$, which uses a
similar scaling relation between the significance, $\zeta$, and
$M_{200a}$. This value also includes a 6\% reduction to the published
mass value to reflect the change in redshift from their photometric
estimate ($z=0.78$) to the true spectroscopic value ($z=0.87$).

\subsubsection{Stellar Mass from SED Fitting}
\label{sec:SED}

\begin{figure}
\centerline{\includegraphics[width=3.7in]{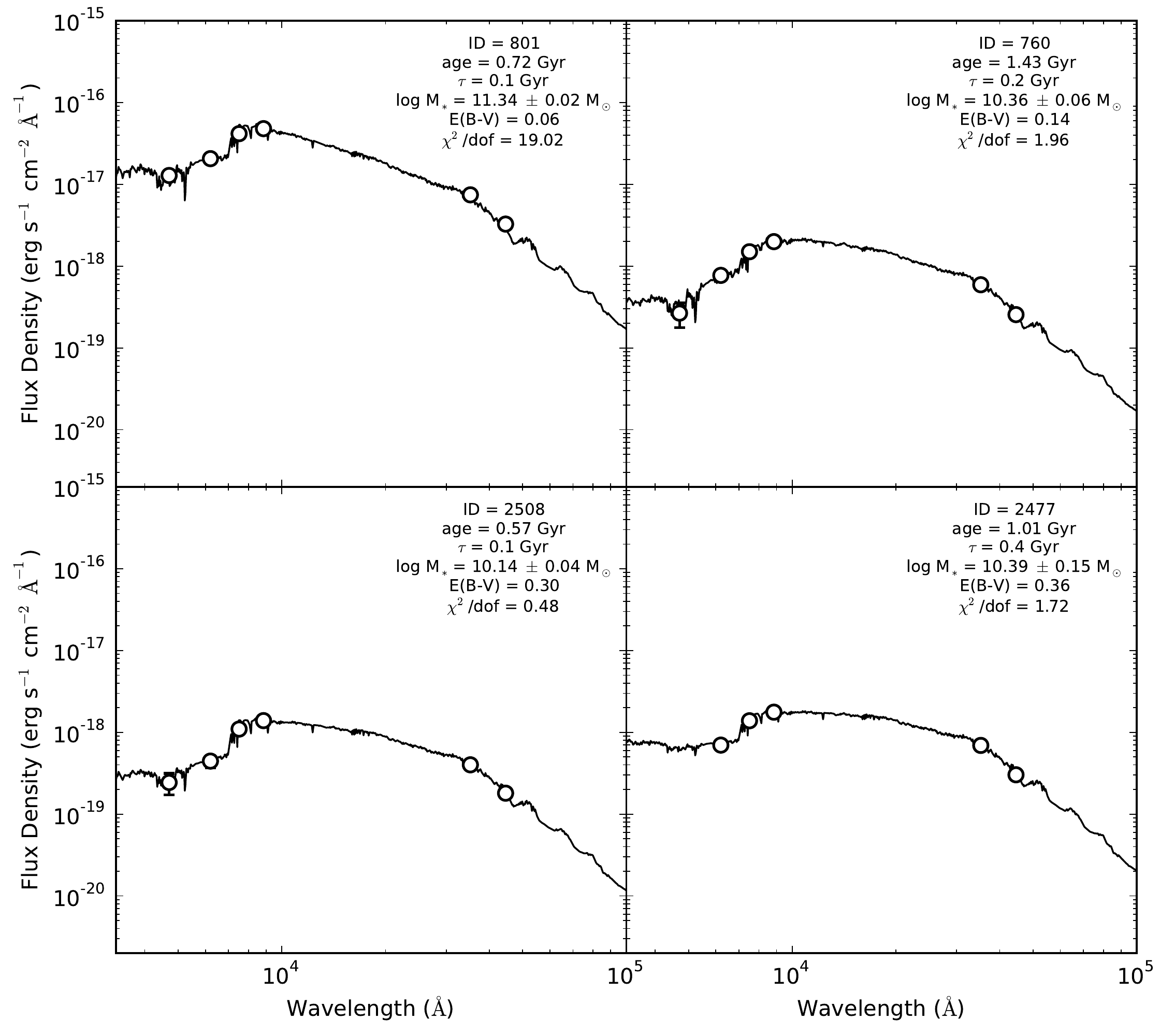}}
\caption{Examples of the SED fitting for four galaxy cluster members
  in \J0102. The solid line represents the best-fit SED for the
  broad-band $g,r,i,z$, \3p6 and \4p5 photometry shown as open
  symbols.}
\label{fig:SEDFits}
\end{figure}

In order to measure stellar population parameters and the stellar mass
content of the cluster using galaxy members, we combine the optical
photometry described in Section~\ref{sec:optical} with the IRAC
photometry described in Section~\ref{sec:irac} by cross matching the
catalogs using a $1.\!\!\arcsec2$ matching radius. We adopt the SExtractor
MAG\_AUTO measurements as estimates of total galaxy magnitudes in the
optical bands, and the $4\arcsec$ diameter aperture corrected
magnitudes as measurements of total flux in the IRAC bands.

We fit the broadband Spectral Energy Distributions (SED) using a
similar methodology to that described in \citet[][see also
  \citealt{Hilton2010}]{Shapley2005}. We augmented the sample of 89
spectroscopic members to a total of 411 selected using photometric
redshifts \citep{BPZ} re-calibrated with the 517 redshifts from seven
clusters in the range $0.4<z<0.9$ from \cite{Sifon2012} and with the
same $griz$ photometry as \J0102. The improved photometric redshifts
had typical $\delta z = 0.04$ in the redshift range $0.8<z<0.9$. We
select galaxies with photometric redshifts within $\Delta z \pm0.06$
of the cluster mean redshift.

We construct a grid of \citet{BC03} solar metallicity models with
exponentially declining star formation histories with 20 values of
$\tau$ in the range $0.1-20$~Gyr, and 53 ages in the range
$0.001-7.0$~Gyr. We adopt a \citet{Chabrier2003} initial mass function
(IMF), and model the effect of dust extinction using the
\citet{Calzetti2000} law, allowing $E(B-V)$ to vary in the range
$0.0-0.5$ in steps of 0.02. The SEDs are fitted by analytically
calculating the normalization $N$ for each model SED where
$d\chi^2/dN=0$, adopting the model with the lowest $\chi^2$ value as
the best fit. The stellar mass $M_*$ is then estimated from the value
of $N$. We therefore fit for a total of four parameters (age, $\tau$,
$E(B-V)$, $M_*$). Errors on the SED parameters were estimated using
Monte-Carlo simulations. Figure~\ref{fig:SEDFits} shows some examples
of fitted SEDs.

Given the small number of SED points compared to the number of free
parameters, coupled with degeneracies between several parameters (age,
$\tau$, $E(B-V)$; see the discussion in \citealt{Shapley2005}), we do
not expect the values of these parameters to be very robust.  However,
stellar mass is well constrained by the IRAC photometry, which probes
the rest-frame near-IR at the cluster redshift, and stellar mass
estimates have been shown to be fairly insensitive to variations in
other model parameters, such as the assumed star formation history
\citep[e.g.,][]{Forster2004, Shapley2005}. We note that uncertainties
in the IMF and the modeling of thermally pulsating AGB stars
\citep[e.g.,][]{Maraston2005, Conroy2009} is likely to lead to the
stellar mass estimates only being accurate to within a factor of two;
these systematic uncertainties are not taken into account in quoted
errors on $M_*$. The total stellar mass for all cluster members within
$r_{200} = 2111\,h_{70}^{-1}$~kpc is $M_{200}^* = 1.31\pm 0.26 \times
10^{13}M_{\odot}$. This suggests a ratio of stellar mass to total mass
of $<1\%$ within $r_{200}$ for \J0102.

\subsection{Cluster Structure}
\label{sec:structure}

\renewcommand{\arraystretch}{1.34} 
\begin{deluxetable*}{ l c c c c c }
\tablecaption{Dynamical Mass for \J0102 components}
\tablehead{
\colhead{Group}    &
\colhead{$N_{\rm gal}$}        &
\colhead{redshift} &
\colhead{$\sigma_{\rm gal}$} &
\colhead{$r_{200}$} &
\colhead{$M_{200a}$}  \\
\colhead{} &
\colhead{} &
\colhead{} &
\colhead{(km~s$^{-1}$)}  &
\colhead{(kpc~$h_{70}^{-1}$)} &
\colhead{$(10^{15}h_{70}^{-1}M_\odot)$} 
}
\hline
\startdata
ACT-CL~J0102$-$4915~(NW)   & 51 & $0.86849 \pm 0.00020$ & $1290 \pm 134$ & $2062\pm233$ &  $1.76_{-0.58}^{+0.62}$ \\
ACT-CL~J0102$-$4915~(SE)   & 36 & $0.87175 \pm 0.00024$ & $1089 \pm 200$ & $1743\pm336$ &  $1.06_{-0.59}^{+0.64}$ \\
ACT-CL~J0102$-$4915~(total)& 89 & $0.87008 \pm 0.00010$ & $1321 \pm 106$ & $2111\pm189$ &  $1.86_{-0.49}^{+0.54}$  \\
\enddata
\label{tab:groups}
\end{deluxetable*}
\renewcommand{\arraystretch}{1.0} 

\begin{figure*}
\centerline{
\includegraphics[width=3.3in]{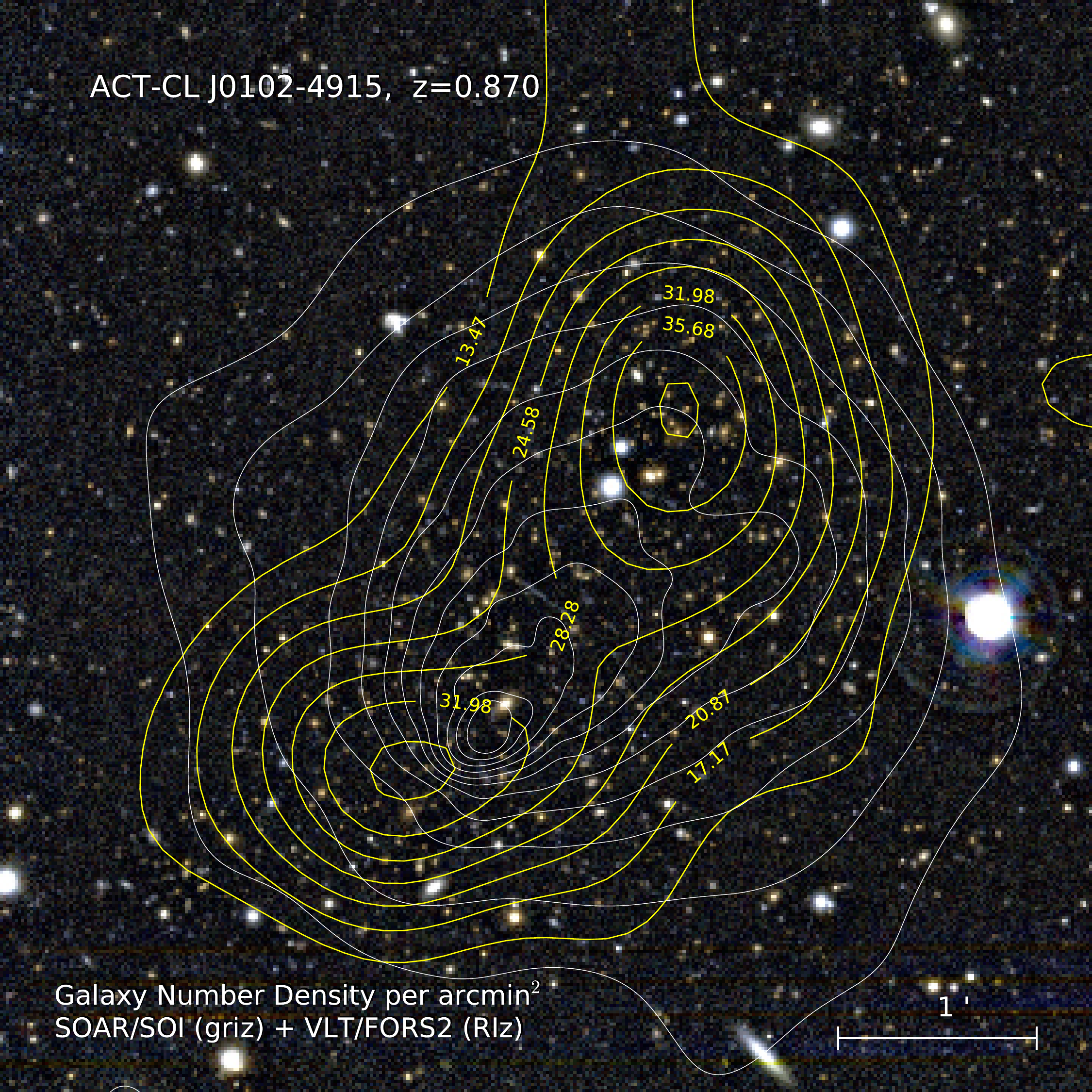}
\includegraphics[width=3.3in]{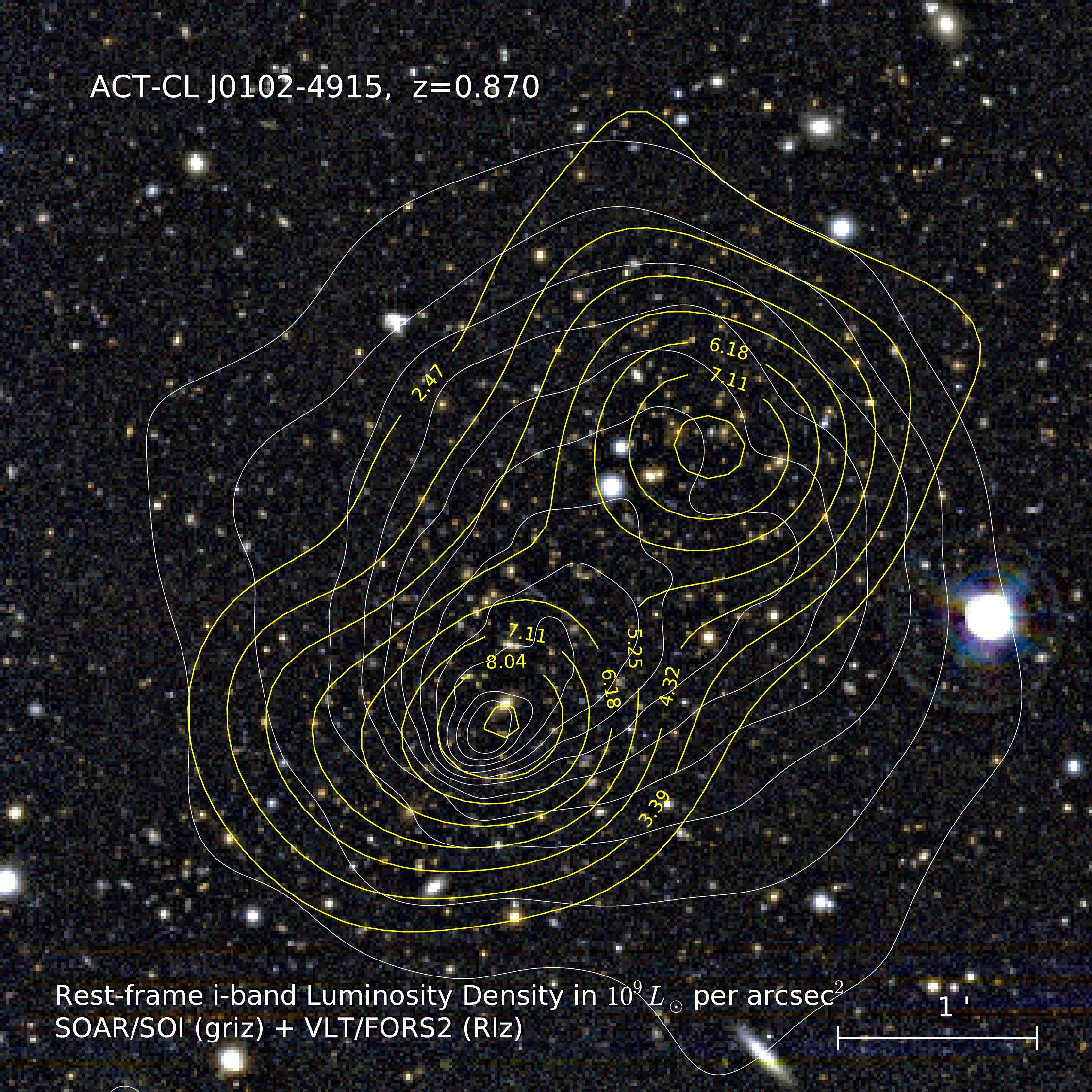}}
\centerline{
\includegraphics[width=3.3in]{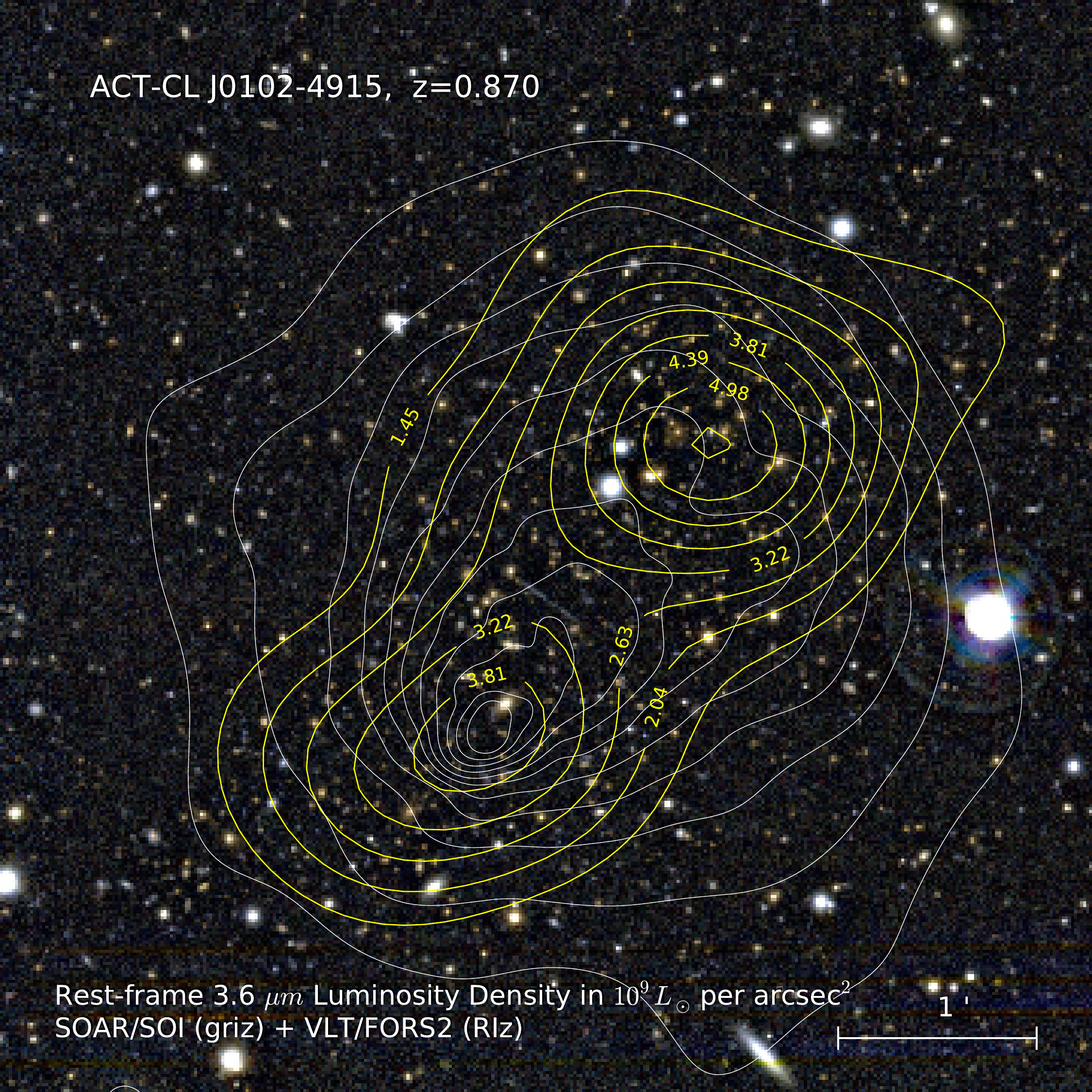}
\includegraphics[width=3.3in]{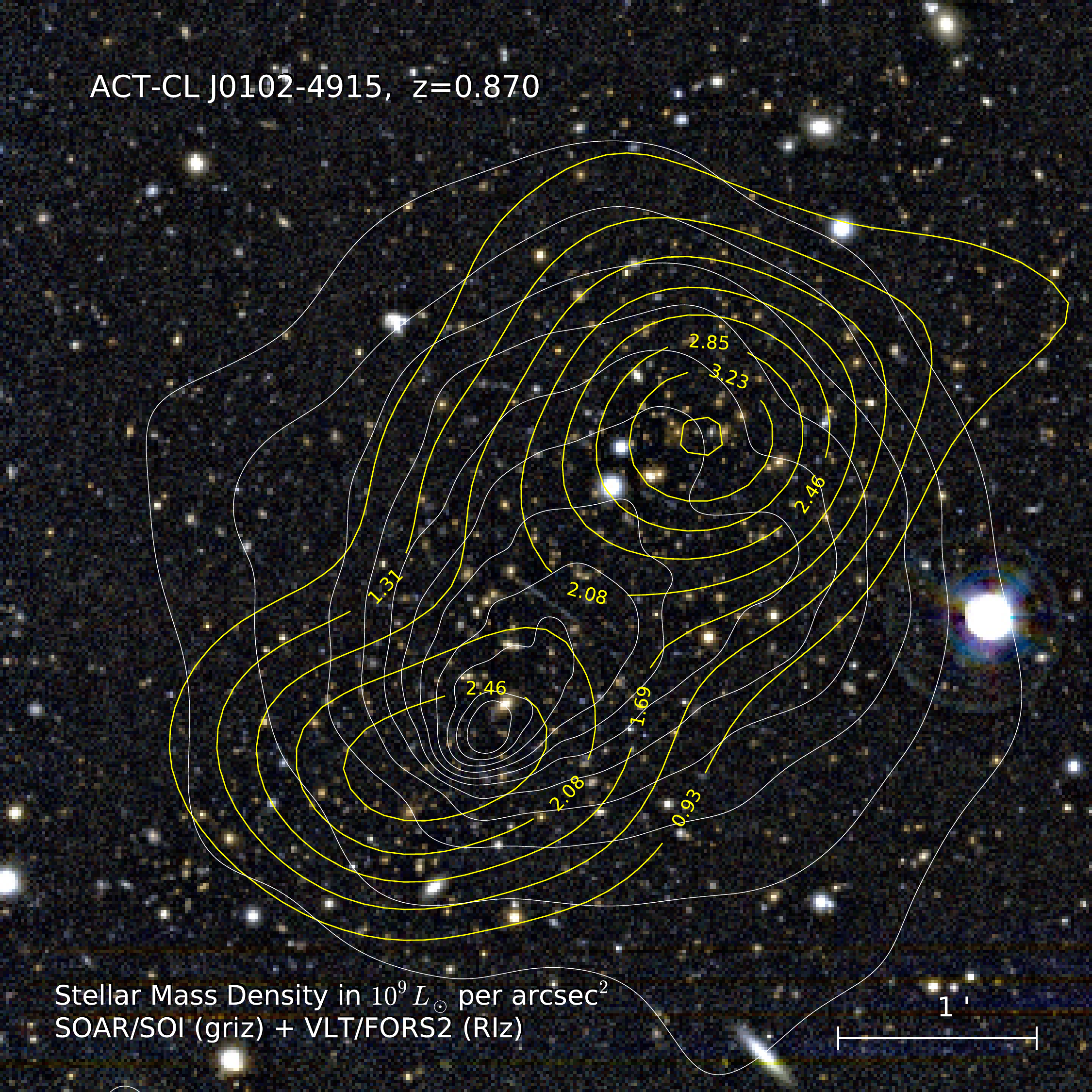}}
\caption{The isopleths from the joint spectroscopically and
  photometrically selected galaxies in \J0102 using four methods. The
  isopleth contours using {\it (upper left)} the number density of
  galaxies per square arcminute, {\it (upper right)} the rest-frame
  optical $i$-band luminosity, {\it(lower left)} the rest-frame
  $3.6\mu$m \spitzer/IRAC luminosity and {\it(lower right)} the
  SED-fitted stellar mass. The luminosity contours are both in units
  of $10^{9}L_\odot$ arcsec$^{-2}$, and the stellar mass map
  is in units of $10^{9}M_\odot$ arcsec$^{-2}$.  The X-ray
  emission is shown as white contours using the same levels from
  Figure~\ref{fig:cplates}. All panels agree on the bimodal structure
  of the galaxy distribution; moreover, both the galaxy number and
  stellar mass density maps have peaks with statistically significant
  offsets with respect to the X-ray emission peak.}
\label{fig:isopleths}
\end{figure*}

\begin{figure}
\centerline{\includegraphics[width=3.9in]{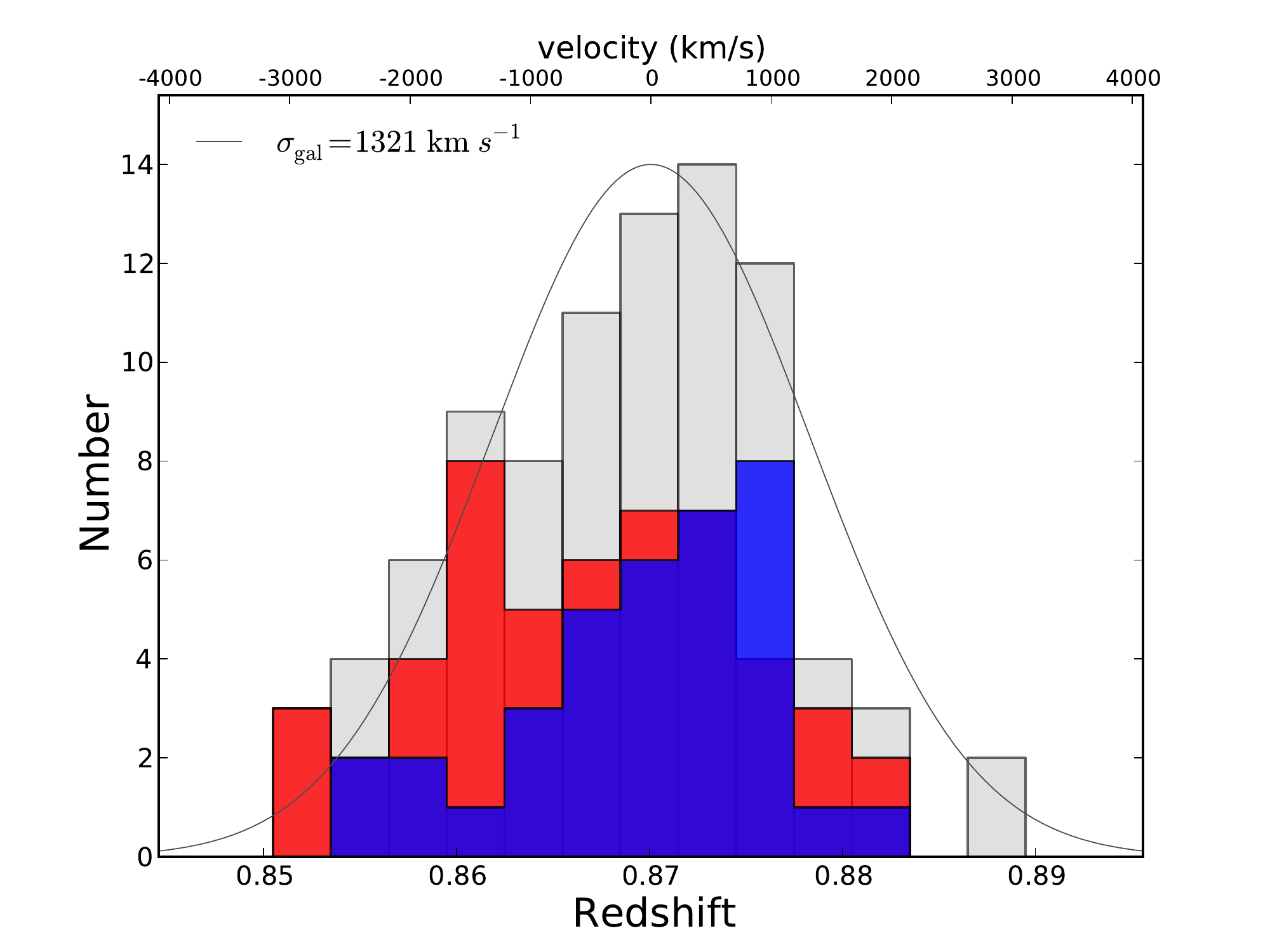}}
\caption{Redshift distributions for spectroscopic member galaxies in
  the NW (red) and SE (blue) clusters subcomponent. The underlying
  distribution for all of the 89 cluster member galaxies for \J0102\ is
  shown in gray.}
\label{fig:zhist_groups}
\end{figure}

In the $\Lambda$CDM paradigm, the dark matter distribution
should be aligned with the distribution of galaxies if the dark
matter particles are indeed collisionless.
The cometary and bullet-like X-ray structure of \J0102\ motivates our
search for a spatial offset between the galaxies (that should
primarily trace the dark matter distribution) and the X-ray gas in a cluster
subcomponent as seen in, for example, \bullet\ \citep{Clowe2004},
\ms1054 \citep{Jeltema2001}, and MACS~J0025.4$-$1222
\citep{Bradac2008}.

Our VLT campaign \citep{Sifon2012} found no clear evidence for
velocity substructure or a bimodal velocity distribution in
\J0102. However, visual inspection of the optical images immediately
suggested an elongated structure in the distribution of the red
galaxies along the NW-SE axis, coincident with the direction of the
structure seen in the X-ray emission (see Figure~\ref{fig:cplates}
left panels).  We investigate the spatial cluster structure in more
detail by constructing four maps: a) the surface density of galaxies,
b) the $i-$band luminosity density, c) the \3p6/IRAC rest-frame
luminosity density and d) the SED-fitted stellar mass density for
spectroscopically and photometrically selected cluster members as
described in Section~\ref{sec:SED}.  These four maps use a $6''$/pixel
grid smoothed with a Gaussian kernel of width $20''$.
Figure~\ref{fig:isopleths} shows our deep optical multi-band imaging
of \J0102\ overplotted with yellow contours of the four density maps
(one per figure panel) along with the same X-ray surface brightness
contours from Figure~\ref{fig:cplates} displayed in white.

In all cases the galaxy distribution is obviously double peaked and a
bright cD-type galaxy sits near each peak.  For the galaxy density
(Figure~\ref{fig:isopleths} top left) and stellar mass density (lower
right) maps, the peak of the X-ray emission lies between the peaks of
the density.  The luminosity densities for the $i-$ and \3p6$-$bands
(Figure~\ref{fig:isopleths} upper right and lower left panels,
respectively) are also double-peaked distributions. The $i$-band
luminosity density, which probes the ultraviolet at $z=0.87$, is
somewhat unusual because it is dominated by the light contribution of
the ultra-luminous Brightest Cluster Galaxy (BCG), discussed further
below. As a result, the SE peak has the higher amplitude in the
$i$-band, and coincides with the peak of the X-ray emission and with
the location of the BCG. In the \3p6 band, which better traces the
stellar mass, the NW peak is dominant, as with the galaxy number
distribution.  Note that the spatial distribution of the
spectroscopically-confirmed cluster members is also double-peaked,
similar to the contours in the top left panel of
Figure~\ref{fig:isopleths}.

Between and nearly equidistant from the two peaks in all four
distributions lies a 20$^{\prime\prime}$-long straight blue arc,
presumably the strongly lensed image of a high-redshift background
galaxy (see black and white insert in Figure~\ref{fig:cplates}). This
feature appears in the trough between the two peaks of all four
distributions.

Regardless of whether we inspect the individual galaxies or their
integrated light contribution, it is clear that there is a double peak
structure in their spatial distribution with nearly parallel contours
that also align with the direction of the lensing arc.
We use this feature to define two components in the cluster (which we
refer to as the NW and SE components) and investigate their masses and
differences in velocities of the components.  Specifically, we divide
cluster galaxies into two groups along the line defined by the pair of
RA and Dec positions (01:03:22.0, $-$49:12:32.9) and (01:02:35.1,
$-$49:18:09.8).  
The velocity dispersions of the NW and SE components are $1290\pm134$
and $1089\pm200$~km~s$^{-1}$, respectively.  We compute their masses
using the same method described in section~\ref{sec:dmass} and obtain
$1.76_{-0.58}^{+0.62}\times 10^{15}h_{70}^{-1}M_{\odot}$ and
$1.06_{-0.59}^{+0.64}\times10^{15}h_{70}^{-1}M_{\odot}$,
respectively. Formally this implies a range of possible mass ratios
from approximately 5:1 to 1:1 between the cluster components; in
subsequent analysis we will consider a 2:1 mass ratio.
Table~\ref{tab:groups} gives computed values for the cluster
components, as well as the cluster total.  The sum of the masses of
the two components determined separately, $(2.8 \pm 0.9)\times
10^{15}h_{70}^{-1}M_{\odot}$, is about 1-$\sigma$ higher than the
total cluster mass inferred from the velocity dispersion of all
galaxies.  We discuss this more in the next section.

The SE component has a line-of-sight peculiar velocity
\footnote{This is estimated as $\Delta v = c(z_1 - z_2) / (1 + z_1)$,
  the peculiar velocity of the SE component in the reference frame of the
  NW component. }
of $586\pm96$~\kms\ with respect to the main (NW) component.  We also
note that the cluster BCG ($z=0.87014\pm0.00030$), which is spatially
located in the smaller SE mass component, has a peculiar velocity
of $731\pm 66$~\kms\ with respect to the main (NW) component.
In Figure~\ref{fig:zhist_groups} we compare the distribution of the
velocities for the NW and SE components shown in red and blue
respectively against the distribution for the full sample. 

We also use the cluster spatial information to obtain the stellar mass
of the components with their respective $r_{200}$ value using the SED
fits from section~\ref{sec:SED}. Following the same component
segregration as described above, we obtain stellar masses of
$(7.5\pm1.4)\times10^{12}\,M_\odot$ and
$(5.6\pm1.3)\times10^{12}\,M_\odot$ 
for the NW and SE components respectively.

\subsection{Combined Mass}

\renewcommand{\arraystretch}{1.55} 
\begin{deluxetable}{c c c c}
\tablecaption{Mass measurements for \J0102.}
\tablehead{
\colhead{Mass Proxy} &
\colhead{Measurement} &
\colhead{Scaling Law} &
\colhead{$M_{200a}$} \\
\colhead{} &
\colhead{} &
\colhead{} &
\colhead{$(10^{15}h_{70}^{-1}M_\odot)$} 
}
\hline
\startdata
${\bm\sigma}_{{\bf gal}}$       & $1321\pm106$~km s$^{-1}$   & $M_{200} - \sigma$    &$1.86_{-0.49}^{+0.54}$\\ 
$T_X$                         & $14.5\pm1.0$~keV          & $M_{500c} - T_{X}$      &$2.58_{-0.59}^{+0.84}$\\
$L_X$                         & $2.19\pm0.11$              & $M_{500c} - L_{X}$      &$2.60_{-0.87}^{+1.36}$\\ 
\bf{\em{Y}}$_{\mbox{\bf{\em{X}}}}$ & $3.20\pm0.24$             & $M_{500c} - Y_{X}$      &$2.88_{-0.55}^{+0.78}$\\ 
$M_{\rm gas}$                    & $2.2\pm0.1 \times 10^{14}\, M_\odot$  & $M_{500c} - M_{\rm gas}$ &$3.10_{-0.66}^{+0.92}$\\ 
\bf{\em{yT}}$_{\bf{CMB}}$        & $490\pm60\,\mu$K          & $M_{200} - y$         &$1.64_{-0.42}^{+0.62}$\\
\\
\hline\\
{\bf Combined}  & \multicolumn{2}{c}{}             &{$\bf 2.16 \pm 0.32 $}\\
\enddata
\tablecomments{$M_{500c}$ masses from $L_X$ and $T_X$ were scaled to
  $M_{200a}$ masses assuming an NFW density profile and the
  mass-concentration relation of \cite{Duffy2008} as described in the
  text. $L_X$ is in units of $10^{45}\,h_{70}^{-2}$erg~s$^{-1}$ and is
  in the 0.5--2.0 keV band. $Y_X$ is in units of $10^{15}\,M_\odot\, {\rm keV}$.
  The combined mass is the minimum $\chi^2$ of the mass proxies
  in bold ($\sigma_{\rm gal}$, $Y_X$, and $yT_{\rm CMB}$). 
}
\label{tab:mass}
\end{deluxetable}
\renewcommand{\arraystretch}{1.}

Table~\ref{tab:mass} compiles our mass estimates for \J0102 from
X-ray, SZ, and optical data sets. While central values range over
nearly a factor of two, all estimates are statistically consistent.
This can be seen by noting that the 2-$\sigma$ mass ranges for all six
mass estimates overlap for the mass range $1.78\times
10^{15}h_{70}^{-1}M_\odot$ and $2.88\times 10^{15}h_{70}^{-1}M_\odot$;
in other words, for any mass in this range, none of the six mass
proxies are more than 2-$\sigma$ away.  For a combined mass estimate
on which to base further conclusions, we construct a $\chi^2$ curve
for a fit to a single combined mass value using the several mass
estimates with their appropriate plus and minus error bars.  For the
three independent mass estimators from velocity dispersion, $Y_X$, and
Sunyaev-Zeldovich decrement, we obtain a minimum $\chi^2$ of 2.7 for 2
degrees of freedom (statistically acceptable) at a mass value of
$M_{200a}=(2.16\pm0.32)\times 10^{15}h_{70}^{-1}M_\odot$.  We do not
combine all six mass proxies (which gives a somewhat higher mass
value), since the four derived from the same X-ray data are all likely
correlated to a degree that is difficult to evaluate.

The above error range includes uncertainties from measurement
error, mass scaling relations, and extrapolation to a common
$M_{200a}$ mass scale. 
Estimates of galaxy cluster masses also may be impacted by a number of
potential systematic errors. Since cosmological conclusions are
strongly sensitive to the estimated cluster mass, it is important to
have a realistic appraisal of these errors.  Here we describe
several relevant sources of systematic error or biases in both measured mass
estimates and in the theoretical comparison with cosmological models.

Dynamical mass estimates are based on the assumption that the system
is dynamically relaxed, which is almost surely not the case in \J0102,
based on the observed spatial galaxy distribution. If we are observing
a merger with the component velocities primarily in the plane of the
sky, then significant kinetic energy can be contained in motions
transverse to the line of sight which are not observed in galaxy
redshifts; the dynamically inferred mass would be lower than the
actual system mass.  This seems to be consistent with the measurements
of the dynamical masses of the two subcomponents in \J0102\ (explored
in Section~\ref{sec:structure}), which when summed, $(2.8 \pm
0.9)\times 10^{15}h_{70}^{-1}M_{\odot}$, is in better agreement with
the X-ray--derived masses than the full galaxy sample.  Therefore the
dynamical mass of the full cluster quoted in Table~\ref{tab:mass} 
may be an underestimate.

A number of groups \citep[e.g.,][]{Ricker2001,Randall2002} have shown
that cluster mergers can result in a transient boost in X-ray
temperaure and luminosity during a merger, compared to the X-ray
emission of the merged cluster once it relaxes. The effect that such
boosts have on the X-ray mass proxies has been more recently
investigated by \cite{KVN}, \cite{Poole2007} and \cite{Yang2010},
among others. \cite{V09} argue that masses estimated from $T_X$ should
be {\it increased} by a factor of 1.17 (which we do not apply) for
merger systems, to account for the fraction of merger kinetic energy
that remains unthermalized.  \citet{KVN} introduced the $Y_X$ mass
proxy which has been shown to be fairly independent of merger dynamics
\citep[e.g.,][]{KVN,Poole2007}. \J0102\ shows strong evidence for
being in the merger process, particularly the compact region of X-ray
enhancement and the double-peaked sky density distribution of
galaxies.  So in our combined mass estimate we strongly prefer to use
$Y_X$, but we note that the masses for all the X-ray proxies are
consistent with each other.  In particular the agreement of our masses
from the $Y_X$ and $M_{\rm gas}$ proxies suggest that our X-ray masses
are not significantly boosted by the merger.

The accuracy of the mass scaling law for the SZ proxy (here $yT_{\rm
  CMB}$) is yet to be determined and in fact is one goal of the ACT
follow-up program.  Still it is interesting to note that the inferred
SZ mass for \J0102\ is only 60\% of the $Y_X$-inferred mass, albeit
consistent within the large errors. The excellent agreement we find
between the predicted SZ signal from the \chandra\ and ACT data (as
illustrated in Figure~\ref{fig:sz-cxo}) suggests that we have a
realistic gas model for the cluster and thus points to the mass
scaling law, particularly at high redshift, as the main source of
possible discrepancy.  A similar difference between the SZ and X-ray
derived masses is also seen for the massive cluster \SPT2106\ at
$z=1.132$ \citep{Foley2011}, while \cite{Andersson2010} find that
SZ-derived masses are some 22\% less than their X-ray masses.
Given six mass proxies for  \J0102\ which are all statistically consistent,
it seems unlikely that our combined mass estimate is subject
to significant systematic error. 

For cosmological model comparisons, the theoretical mass function for
clusters this massive is not precisely determined. In principle,
direct numerical simulations of dark matter cosmology in large volumes
and low mass resolution can determine the high end of the mass
function in a straightforward way. The largest simulations to date
have comoving volumes of roughly 2000 Gpc$^3$, on the order of our
Hubble volume, which nonetheless contain perhaps only a single cluster
with mass $10^{15}\,M_\odot$ at $z=1$.  An analytical model of the
high-mass mass function accurate at the 10\% level has not yet been
demonstrated, so the mass functions used for comparison involve
extrapolation from lower-mass halos where statistics from numerical
simulations are better. Care also must be taken to ensure that initial
conditions in simulations are sufficiently accurate; the commonly used
Zeldovich Approximation for initial conditions may underestimate the
number of high-mass haloes by 10\% to 30\% compared to more accurate
second-order Lagrangian perturbation theory
\citep{Crocce2006,Berlind2010}. Finally, large-volume simulations give
dark matter halo masses, which then must be translated into total halo
masses, with possible systematic departures from the commonly assumed
universal baryon to dark matter ratio in large clusters.

\subsection{Cluster Evolution}
\subsubsection{Color-Magnitude relations}
\label{sec:cmd}

\begin{figure}
\centerline{
\includegraphics[width=3.6in]{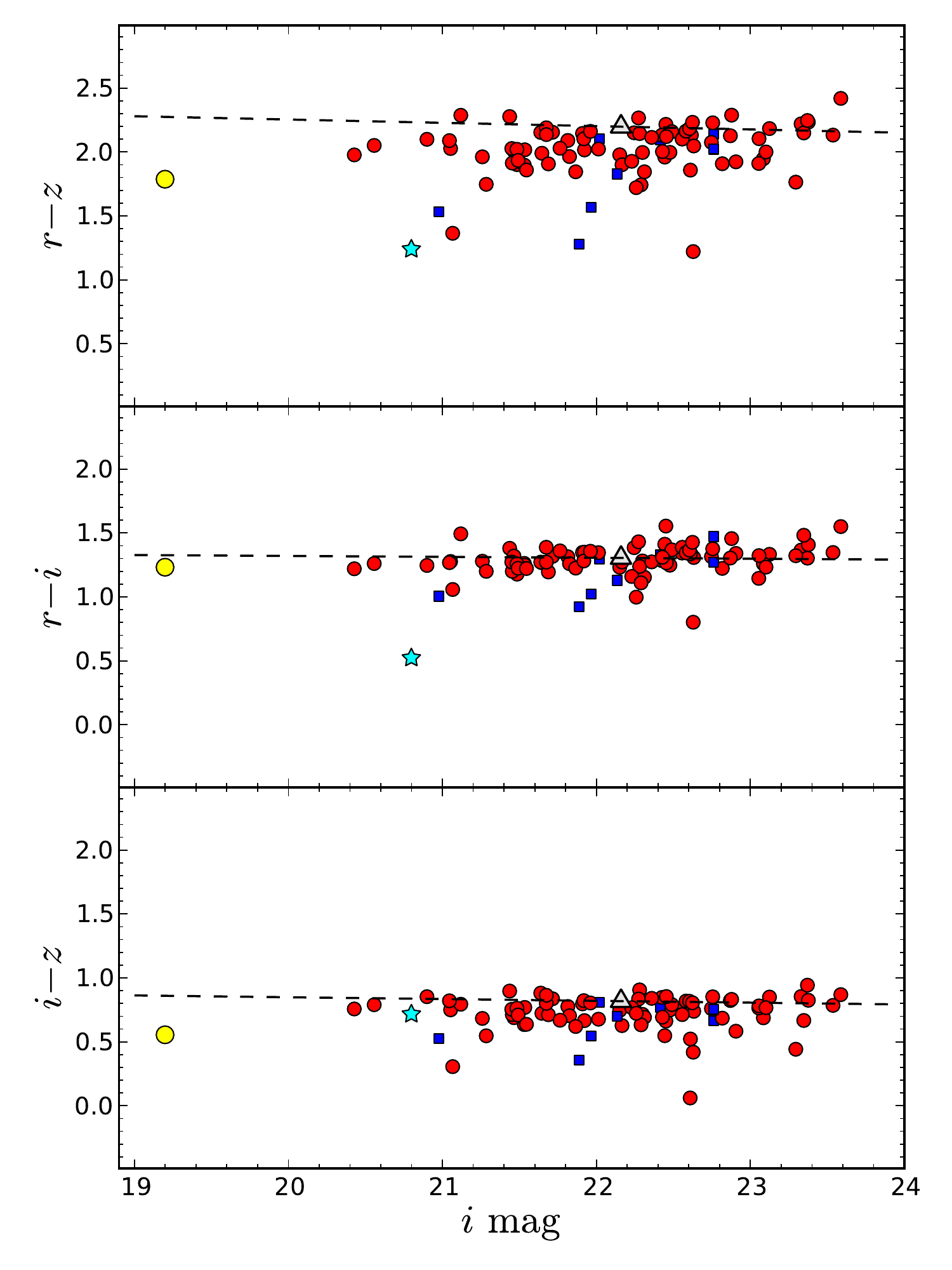}
}
\caption{The optical color-magnitude relations for \J0102 for
  spectroscopically confirmed cluster members. Absorption and emission
  line galaxies are represented by red circle and blue squares
  respectively while the BCG is denoted by a yellow circle. The one
  spectroscopically confirmed AGN is represented by a cyan star. The
  dashed lines represent the fitted relations for the cluster
  RXJ0152$-$1357 at $z=0.83$ from \cite{Blakeslee2006} for the
  corresponding HST/ACS filters redshifted to $z=0.87$. The gray triangle
  represents the observered magnitude of a $M^*$ galaxy ($i=22.16$) in
  the CMR.  }
\label{fig:cmd1}
\end{figure}

\begin{figure}
\centerline{
\includegraphics[width=3.6in]{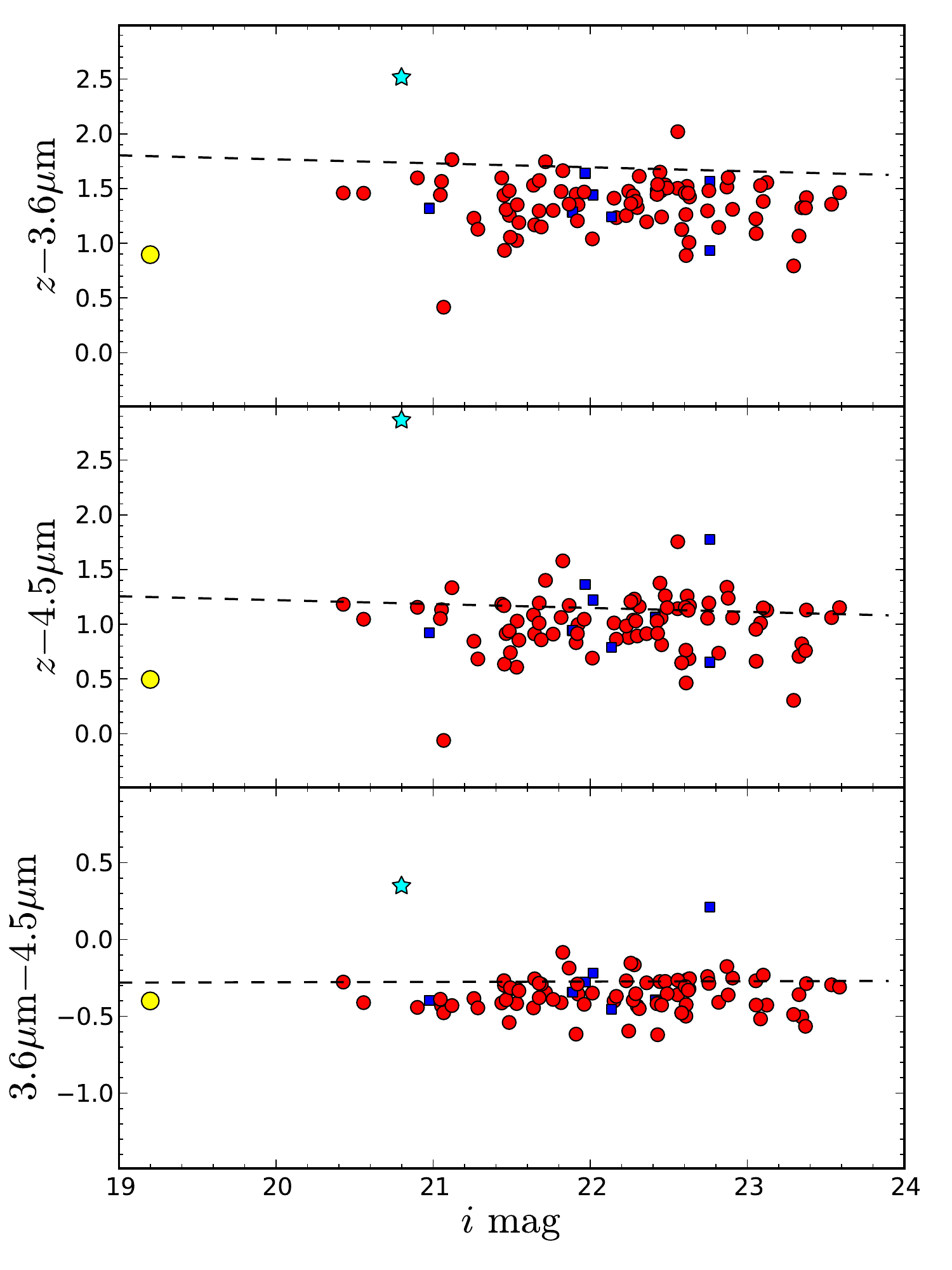}
}
\caption{The optical to infrared color-magnitude diagram for cluster
  member galaxies. Symbols are as in Fig.~\ref{fig:cmd1}. The dashed lines
  again represent the fitted relations for the cluster RXJ0152$-$1357 at
  $z=0.83$ from \cite{Blakeslee2006}, now transformed to the \spitzer/IRAC
  bands as described in the text.}
\label{fig:cmd2}
\end{figure}

Considerable effort has been devoted to understanding the formation
and evolution of ellipticals in clusters up to $z\simeq1.4$
\citep[e.g.,][]{Lidman2008, Hilton-09}. A consistent picture has
emerged in which these galaxies are a well-behaved class of objects
with structural and chemical properties obeying simple power-law
scaling relations. The high degree of homogeneity observed in these
relations and in their stellar population suggests formation at high
redshift followed by passive evolution of their galaxies
\citep{Bower1992, Barrientos1996, Ellis1997, Holden2005, Mei2009}. In
principle, when precise photometry is available measurements of the
intrinsic scatter about the color-magnitude relation can be used to
constrain the ages of the constituent stellar populations of cluster
galaxies.

Here we study the optical and optical/IR color-magnitude relations
(CMR) for \J0102\ to assess the evolutionary stage of the cluster. We
construct the color-magnitude diagrams for spectroscopically confirmed
cluster members using the basic approach described in
\citet{Blakeslee2006} and \citet{Menanteau2006}. For total magnitudes
we use the SExtractor MAG\_AUTO values (with a Kron factor of 2.5) and
we measure colors within the galaxy effective radii $R_e$ which were
computed using GALFIT \citep[version 3.0,][]{GALFIT} to fit each
galaxy to a \cite{Sersic1968} model from the PSF-convolved $i-$band
images. Bright nearby objects are fit simultaneously while faint ones
are masked.  Magnitudes in each passband are then measured within a
circular aperture of radius $R_e$ while enforcing a minimum radius of
1.5 pixel.

In Figures~\ref{fig:cmd1} and \ref{fig:cmd2} we show the optical and
optical-to-infrared color-magnitude diagrams for spectroscopically
confirmed cluster members. Absorption and emission line galaxies are
represented by red circles and blue squares respectively while the BCG
is denoted by a larger yellow circle. In order to ease the comparison
between panels all colors are shown with respect to the total $i-$band
magnitude.  In order to compare with the expected colors of passively
evolving system in figure~\ref{fig:cmd1} we show for comparison the
fitted optical CMRs for the cluster \rxj\ at $z=0.83$ from
\cite{Blakeslee2006} which was observed with the corresponding HST/ACS
filter set and redshifted to $z=0.87$ using a solar metallicity
\citet{BC03} burst model with $\tau=1.0$~Gyr formed at $z_f=5$. We
also display for reference the typical magnitude for an $M^*$
($i=22.16$) galaxy at the cluster redshift in the CMR as a gray
triangle. In Figure~\ref{fig:cmd2} we show the fitted $i-z$ vs.~$i$
relation from Fig.~\ref{fig:cmd1} transformed to the \spitzer/IRAC
band using the same stellar population model.

Despite some scatter, a ``red sequence'' of passively
evolved galaxies is quite distinguishable and it is in good agreement
with the observed relations for \rxj. However, unlike
\cite{Blakeslee2006} we see bluer early-type galaxies brighter that
$M^*$ possibly due to recent star-formation triggered by the merger of
the cluster. We note however, that our classification of galaxies is
only based on the color and visual information in our seeing-limited
ground-based imaging.

From the figures is also clear that the BCG is extremely luminous
($\sim1.5$~mag brighter that any other galaxy) and significantly bluer
in its optical colors than expected from a passively evolving
elliptical galaxy. In the following section we describe the BCG
properties in more detail.

\subsubsection{Luminosity of Brightest Central Galaxy}

\begin{figure}
\centerline{
\includegraphics[width=3.9in]{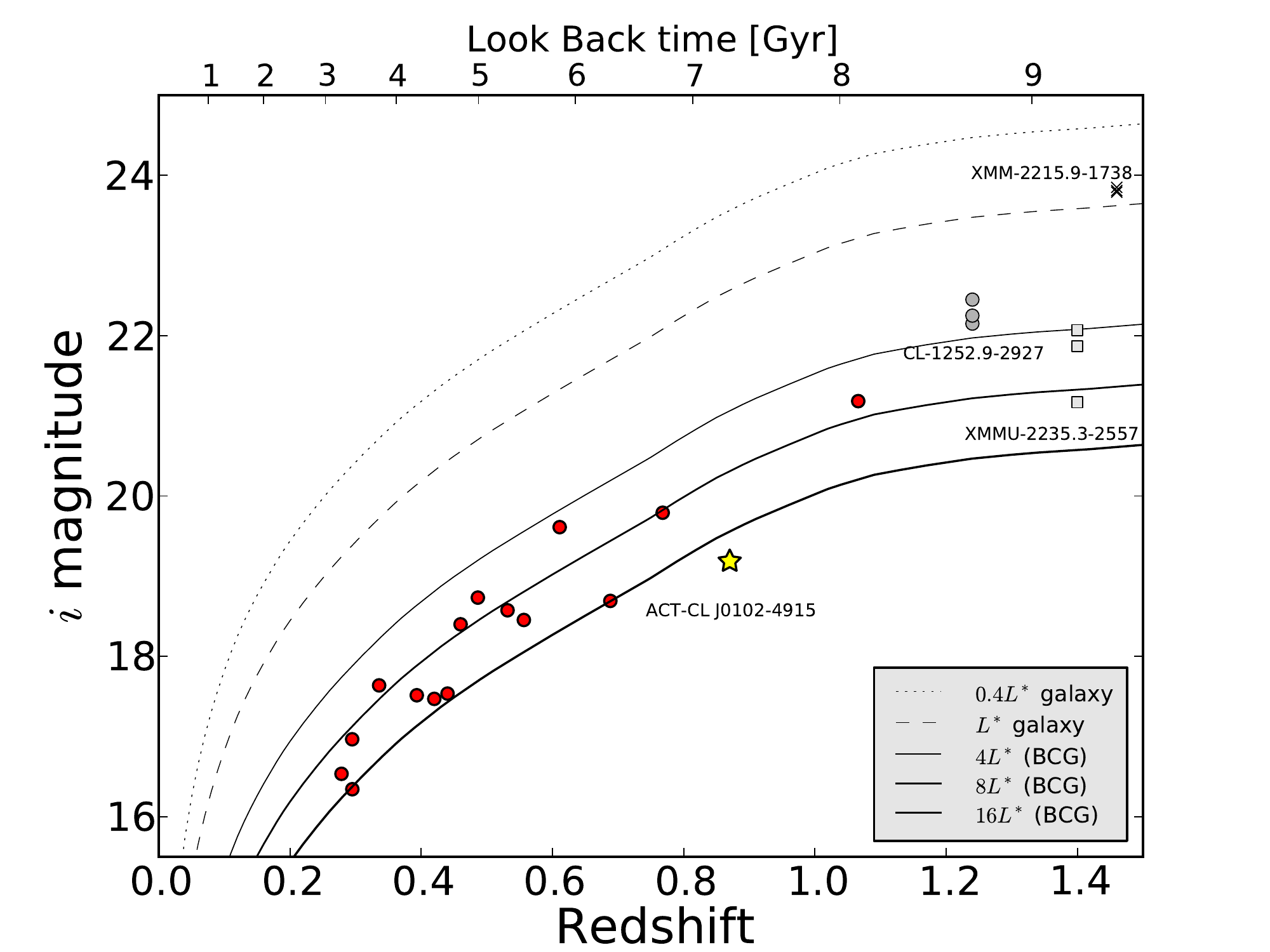}
}
\caption{The observed $i$-band magnitudes for the BCGs in the SZ
  cluster sample reported in \cite{Menanteau-SZ} compared to the BCGs
  of several high redshift clusters. We highlight the BCG of \J0102 as
  a yellow star. For comparison we show the curves for the expected
  $i$-band magnitude as a function of redshift for elliptical galaxies
  over a wide range intrinsic luminosities (i.e. the curves represent the
  observed magnitude for galaxies of the same luminosity).}
\label{fig:BCG}
\end{figure}

In Figure~\ref{fig:BCG}, we plot the observed $i$-band magnitude of
the BCGs of ACT clusters in the southern sample, plus the BCG's for
three X-ray--selected, massive, high redshift clusters:
RDCS~1252.9$-$2927 \citep{Blakeslee2003,Rosati-04, Demarco2007} at $z=1.237$,
XMMU~J2235.3$-$2557 \citep{Mullis-05,Rosati-09} at $z=1.393$ and
XMMXCS~J2215.9$-$1738 \citep{Stanford-06,Hilton-09} at $z=1.457$. For
clarity the BCG for \J0102\ is shown as the yellow star, with a
luminosity of $\simeq22L^*$ in the $i-$band. This is clearly an
extremely luminous object and it is the brightest BCG in the ACT
southern sample.  We also plot the expected apparent $i$-band
magnitudes of elliptical galaxies of different luminosities as a
function of redshift. We use $L^*$ in the $i-$band as defined by
\cite{Blanton2003} at $z=0.1$ ($M^*_i=-20.82 + 5\log(0.7)$) and allow
passive evolution according to a solar metallicity, \citet{BC03}
$\tau=1.0$~Gyr burst model formed at $z_f=5$.  The resulting
magnitude-redshift relation is plotted for a range of luminosities
($0.4L^*$, $L^*$, $4L^*$, $8L^*$ and $16L^*$) aimed at representing
the brightest observed BCGs.

The VLT/FORS2 spectrum of the BCG in \J0102 has the signature of an
E+A+[\ion{O}{2}] galaxy. This galaxy is a possible high-redshift analog
of NGC~1275, the blue and radio-loud cD galaxy in the Perseus cluster
at $z=0.018$ \citep{McNamara1996}. Blue BCGs like this one usually
live in the centers of cool core clusters \citep[e.g.][]{Santos2011}, and in this case the BCG is
located within the cool, low entropy, merging core in \J0102
revealed by the \chandra\ data.
We also obtain the physical size of the BCG from the structural
parameters estimated using GALFIT in section~\ref{sec:cmd}. The galaxy
has an effective radius $R_e=10.4~h_{70}^{-1}$kpc for a S\'ersic index
$n=1.3$. 
It is likely that this BCG is the result of a recent galaxy merger; we
speculate that this extreme galaxy resulted from a
merger of two BCGs during the cluster collision.

\section{Discussion}

\subsection{A New Bullet Cluster?}

Given the absence of a weak-lensing mass reconstruction, we use the galaxy
distribution as a proxy for the total mass distribution.  This should
be a robust approximation since for both 1E0657$-$56 \citep{Randall2008}
and MACS~J005.4$-$1222 \citep{Bradac2008} the galaxy and total
matter distribution closely track each other.
Figure~\ref{fig:isopleths} (upper left panel) shows that the SE galaxy
distribution precedes the peak gas location by
$\simeq22^{\prime\prime}$ ($173\,h_{70}^{-1}$kpc) in the approximate
direction of motion of the merger. This suggests \J0102\ as another
``bullet'' cluster undergoing a high-velocity merger event resulting
in a significant spatial separation in the main baryonic (gas) and
total mass components.  Moreover, \J0102\ is at a significantly
earlier cosmic epoch, a lookback time around 4 Gyr longer than
1E0657$-$56. However, this picture requires confirmation through weak
lensing mass mapping to determine the spatial distribution of the dark
matter.

\subsection{Rarity of \J0102}

\begin{figure}
\centerline{\includegraphics[width=3.9in]{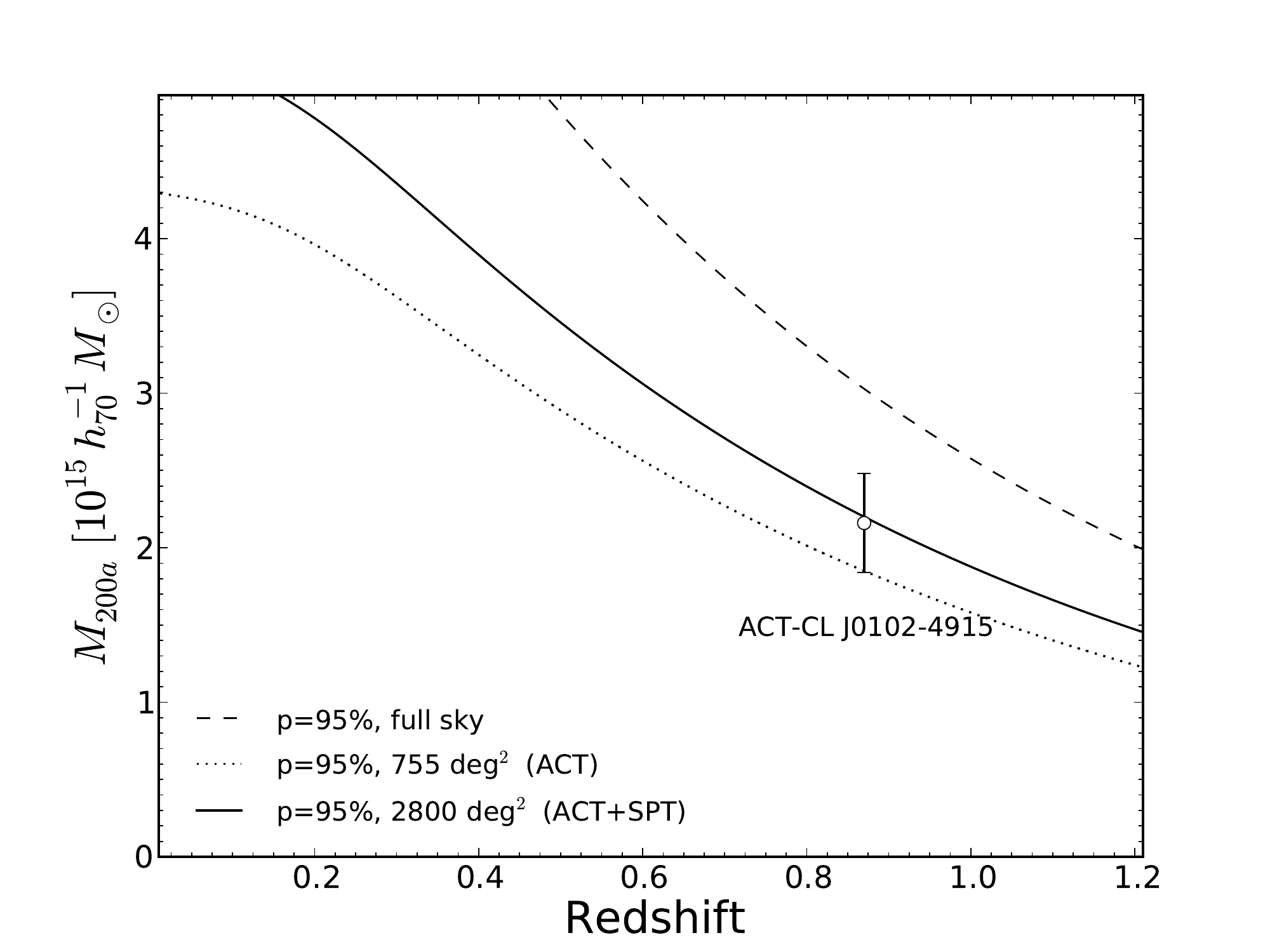}}
\caption{The exclusion curves, $M_{200a}(z)$, from the
  \cite{Mortonson2011} fitting formulas where a single cluster above
  the corresponding curve would conflict with Flat $\Lambda$CDM and
  quintessence at 95\% C.L for both sample and parameters
  variance. The black dashed and dotted lines represent the exclusion
  curves for the full sky and the ACT survey analyzed region of
  755~deg$^2$ (i.e. the mass limit for a cluster at a given redshift
  that it is less than 5\% likely to be found in 95\% of the
  $\Lambda$CDM parameter probability distribution). We also show the
  same exclusion limit for 2800~deg$^2$ (ACT+SPT area) as the black
  solid curve. The open circle shows our combined mass estimate for
  \J0102, with 1-$\sigma$ error bars.}
\label{fig:exc}
\end{figure}

\J0102 is a rare, massive high-redshift cluster.  We use the
convenient fitting formulas provided in appendix~A of
\cite{Mortonson2011} to compute the confidence level at which the
existence of a cluster like \J0102 would be expected in standard
$\Lambda$CDM cosmological models, where bound objects grow via
gravitational instability from Gaussian initial density fluctuations.
In Figure~\ref{fig:exc} we show $M_{200a}(z)$ exclusion curves for
which a single cluster with mass $M_{200a}$ above the corresponding
curve would conflict with $\Lambda$CDM and quintessence at 95\%
confidence level, including both sample and cosmological parameter
variance. In other words, the exclusion curves represent the mass
threshold as a function of redshift for which a cluster is less than
5\% likely to be found in a survey region for 95\% of the $\Lambda$CDM
parameter variance.

For purposes of addressing how rare \J0102 is, in Fig.~\ref{fig:exc}
we plot the exclusion curves for the full sky, the region analyzed for
the ACT survey (755~deg$^2$) where \J0102 was discovered, and the
combined ACT+SPT survey region (2800~deg$^2$ of which 455~deg$^2$ were
observed by both ACT and SPT) as this is the most massive cluster at
$z>0.5$ in the combined area and detected by both experiments. Our
combined estimated 1-$\sigma$ mass range for the cluster (in Table
\ref{tab:mass}) ranges from $1.84$ to
$2.48\times10^{15}\,h_{70}^{-1}M_\odot$.  
At the cluster redshift of 0.87, the lower end of this mass range
falls below the ACT+SPT region exclusion curve, while the upper end of
this range falls well below the full-sky exclusion curve.
In other words, the cluster is not an unlikely occurrence in the
ACT+SPT sky region, provided that its actual mass is 1-$\sigma$ or
more below our nominal mass of $2.16\times10^{15}\,h_{70}^{-1}M_\odot$.
If the cluster mass is 1 to 2-$\sigma$ above this nominal mass, then
it is a rare occurrence in the ACT+SPT sky region but not unexpected
in the entire sky. The cluster mass would have to be about 3-$\sigma$
above the nominal mass for it to lie above the 95\% exclusion curve
for the entire sky. We conclude that, while this cluster is clearly
rare, it is not massive or early enough to put significant pressure
{\it by itself} on the standard cosmological model.

Because the high-mass cluster mass function is steeply falling with
mass, the signals we observe are more likely to be from a cluster with
mass below our nominal mass of $2.16\times 10^{15}h_{70}^{-1}M_\odot$
and a positive measurement error than from a cluster with mass above
this level and a negative measurement error. In other words, the
cluster mass is more likely to be near the lower end of our nominal
mass range than the upper end. \cite{Lima2005} give a discussion of
this Eddington bias and compute a correction for this effect, which in
our case is of order 5\%. This correction does not qualitatively
change any conclusions presented here.

\cite{Lee-Komatsu2010} investigated the rarity of \bullet\ in
cosmological simulations down to $z<0.5$ based on its large merger
velocity and found the existence of such a merging cluster to be in
tension with the expectations of \lcdm. They suggested that such high
merger speeds could potentially be more common at high redshift,
while, on the other hand, massive, $\sim10^{15}M_\odot$ clusters
are rare at earlier times. Using
simulations, \cite{Forero-Romero2010} studied the expected probability
distribution of the displacements between the dark matter and gas
cores, as observed in \bullet, and found it to be expected in 1\%-2\%
of the clusters with masses larger than $10^{14}M_\odot$.

Here we briefly consider the rarity of \J0102\ as a massive cluster
undergoing a major merger with a mass ratio of $\sim$ 2 to 1 at
$z=0.87$, ignoring its merger velocity and spatial separation between
gas and dark matter.  We examined the output of a large-volumne
($3.072\,h^{-1}$ Gpc)$^3$ $\Lambda$CDM cosmological N-body simulation
with 29 billion particles performed by I.~Iliev with the Cubep3m code
at NIC Juelich (G.~Yepes, private communication).  Within this large
volume there is only one cluster of mass
$1.9\times10^{15}\,h_{70}^{-1}M_\odot$ at $z\simeq 1$, which implies a
very low number density of around $10^{-11}$Mpc$^{-3}$ for such
clusters.  This system has only one small potential substructural
feature with a mass of around $10^{13}\,M_\odot$, suggesting a recent
minor merger with mass ratio of 100 to 1.
The comoving volume of the entire sky between $z=0$ and
$z=1$ is $54.7 \,h^{-3}$Gpc$^{3}$ and hence the examined simulation
samples roughly half of the total cosmic volume 
accessible by SZ cluster observations.
We conclude that a merger event of the mass and mass ratio similar to
the one we are witnessing in \J0102\ is a rare event
in the current set of large volume simulations. Significantly larger
simulations, or improved analytic approximations to halo growth
and evolution, will be required to assess directly the
likelihood of massive high-redshift merger events such as \J0102. 

\subsection{Merger Speed}

\begin{figure}
\centerline{
\includegraphics[width=3.3in]{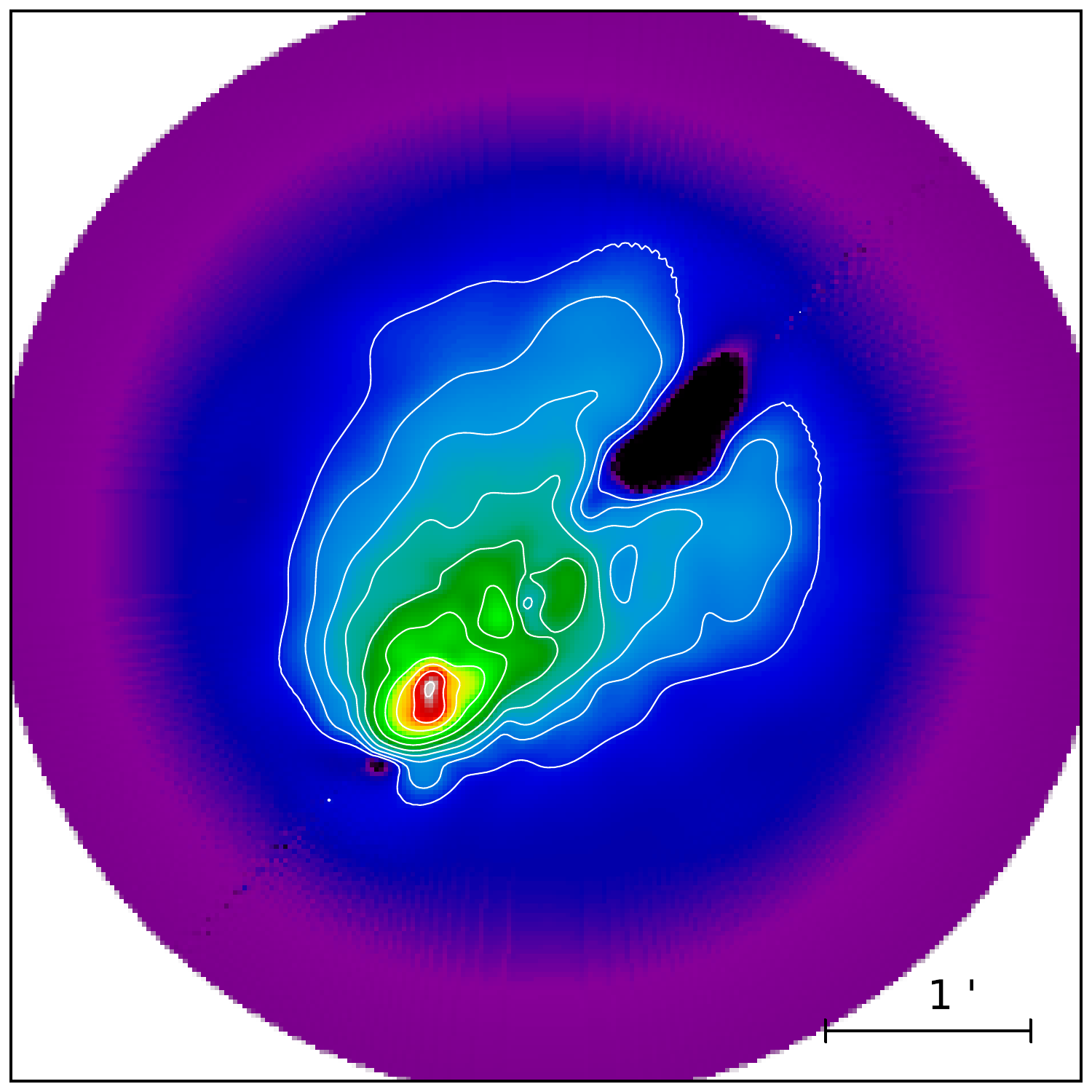}
}
\caption{Map of the electron density in the mid-plane of \J0102\ from
  a deprojection of the \chandra\ image.  The contours are at density
 values of  0.002, 0.003, 0.004, 0.006, 0.008, 0.011, 0.016, 0.023,
 0.032, 0.045 cm$^{-3}$.}
\label{fig:deprojec}
\end{figure}

We have argued above that the depression in the X-ray surface
brightness northwest of the merging bullet is a ``wake'' caused by the
passage of one cluster through the other. In order to study this
feature in more detail, we examine the map of the deprojected electron
density in the cluster mid-plane (i.e., the plane of the sky) of
\J0102\ (see Figure~\ref{fig:deprojec}) .  The merging bullet is
indeed dense with a peak density of $n_e=0.047$ cm$^{-3}$, while the
outskirts fall to a value $n_e\simeq 2\times10^{-4}$ cm$^{-3}$.
Formally our simple deprojection procedure fails in the region of
depressed surface brightness (presumably because the cluster is not
axially symmetric as assumed) and the density values here were pegged
to zero; this area appears black in Fig.~\ref{fig:deprojec}.

Nonetheless, clearly the NW region of depressed
X-ray surface brightness requires a region of extremely low density in
the cluster interior.  The region of depressed density is about
0.8--1$^{\prime}$ ($370h_{70}^{-1}-460h_{70}^{-1}$ kpc) wide. It
begins approximately 2.6$^{\prime}$ ($1200h_{70}^{-1}$ kpc) behind the
leading edge of the bullet toward the northwest and can be traced into the
core of the cluster to within about 1$^{\prime}$ ($460_{70}^{-1}$
kpc) from the bullet. At locations even closer to the bullet, the
deprojected density map shows a complex, disorganized set of peaks and
valleys suggesting the presence of merger-driven turbulence behind the
cold merging core.

The merger speed can be estimated from the peculiar velocity
difference between the SE and NW galaxy concentrations as $v_{\rm
  merg} = v_{\rm pec} / \sin(\theta) = 586 / \sin(\theta)$ km
s$^{-1}$.  The collision is likely taking place close to the plane of
the sky, otherwise the clear morphological features of that merger,
i.e., sharp leading edge, post bullet turbulence, and the wake
(Figure~\ref{fig:deprojec}), would not be so prominent.  Working from
this scenario we assume low inclination angles of $\theta =15^\circ$
and $30^\circ$ that provide merger speeds of $2300$ km s$^{-1}$ and
$1200$ km s$^{-1}$.  An accurate estimate of the merger speed will
require detailed N-body/hydro simulations.

\subsection{Extended Radio Emmission around ACT-CL~J0102$-$4915?}

\begin{deluxetable*}{c c c c c c c c}
\tablecaption{Summary of SUMSS radio source positions and fluxes near \J0102.}
\tablehead{
\colhead{Source} &
\colhead{Name} &
\colhead{RA} &
\colhead{Dec.} &
\colhead{$F_\nu$(mJy)} &
\colhead{size} &
\colhead{P.A.} &
\colhead{in SUMSS} 
}
\hline
\startdata
A  & SUMSS J010246-491435  & 01:02:46.8  & -49:14:35.7 & $18.2 \pm 1.5$ &$70''\!\!\times59''$  & $64^\circ$  & yes \\
B  & SUMSS J010252-491316  & 01:02:52.4  & -49:13:16.7 & $ 7.2 \pm 1.2$ &unresolved & --& no\\          
C  & SUMSS J010301-491705  & 01:03:01.1  & -49:17:05.0 & $ 7.8 \pm 1.2$ &unresolved & --& no\\
\enddata
\tablecomments{Only the first object (A) was cataloged, the latter 2
  sources were analyzed by us using IMFIT (AIPS) on the mosaic
  image. Sources are ordered in increasing right ascension.}
\label{tab:SUMSS}
\end{deluxetable*}

\begin{figure*}
\centerline{
\includegraphics[width=3.5in]{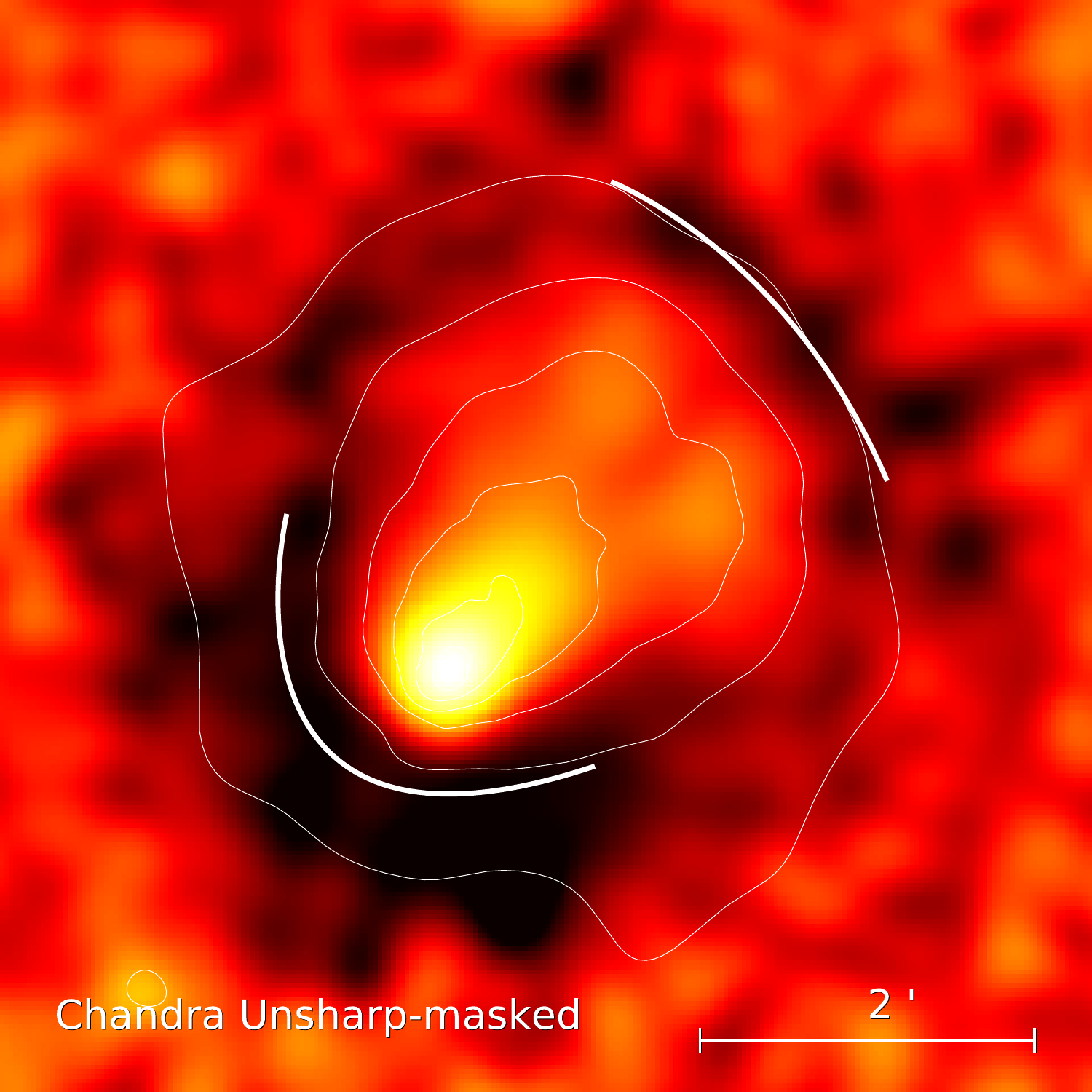}
\includegraphics[width=3.5in]{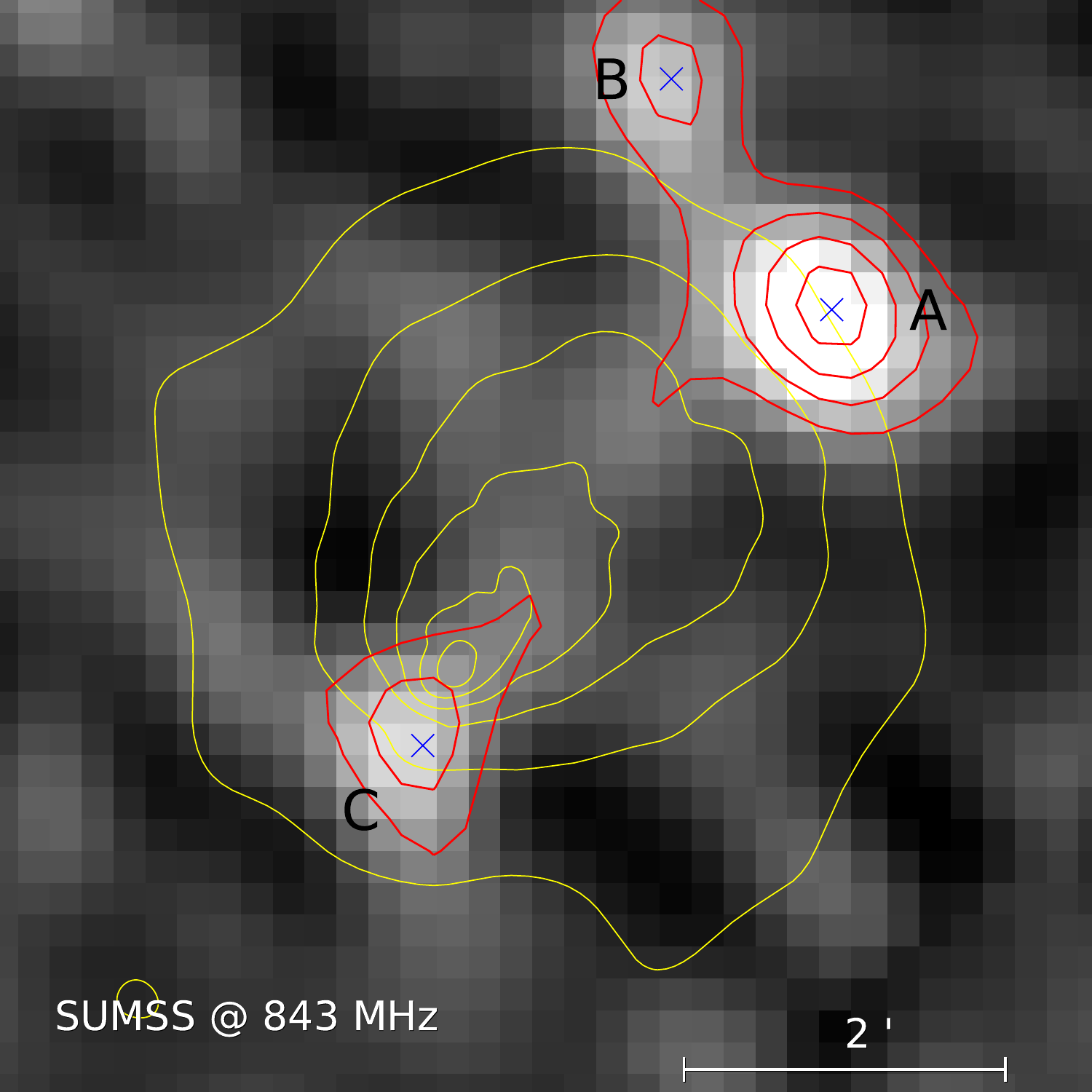}
}
\caption{Unsharp-masked Chandra 0.5--2.0 keV image for \J0102 created
  from the subtraction of two images convolved with Gaussians of
  $\sigma=7\arcsec\!\!.5$ and $30''$ with the high-frequency features
  highlighted as thick white curves ({\em left}). The SUMSS 843~MHz
  image of the same region showing the position of the radio sources
  A, B, and C on the map and their respective contours level are shown
  in red ({\em right}). In both images and for reference
  we overlay the same X-ray contours used in Figure~\ref{fig:regions}}
\label{fig:SUMSS}
\end{figure*}

There is increasing observational evidence \citep[see][for a
  review]{Ferrari2008} that the large amount of energy released
($\sim10^{64}$~erg) in massive cluster mergers can accelerate
particles to relativistic speeds that in the presence of an
intracluster magnetic field can produce synchrotron radiation.  These
diffuse, non-thermal emission features, called radio relics or halos,
have no obvious connection with individual cluster galaxies and have
typical sizes of about 1 Mpc.
We searched for such radio features around \J0102\ using Sydney
University Molonglo Sky Survey (SUMSS) archival data at 843~MHz.  The
radio mosaic at the cluster location has an angular resolution of
$45''\!\!\times60''$ (major axis aligned N-S) and a positional
uncertainty for bright sources ($> 20$ mJy beam$^{-1}$) of $1''-2''$
\citep{Mauch2003}.
In Figure~\ref{fig:SUMSS} (right panel) we show the SUMSS archival map
with the location of three radio sources, labelled A, B and C,
indicated with blue crosses. Only source A was cataloged in
\cite{Mauch2003}; the latter two sources (B and C) were analyzed by us
using IMFIT within the Astronomical Image Processing System (AIPS).
Under the assumption that the typical position error for faint sources
scales as the beam size times the signal-to-noise ratio, we estimate
positional errors for sources B and C of $\sim7''$, which we use in
the search for possible counterparts (galaxies, AGN) in our optical and
X-ray data.
There is an excellent compact optical, IR, and X-ray counterpart for
source B ($<1''$ away), making an AGN identification for this source
likely. 
The nearest X-ray source to position C is $8''$ away which is at the
limit of the radio source positional uncertainty so a clear
identification is not possible.  Furthermore this radio source is
located right at the apex of the merging bullet, at the plausible
location of a bow shock.
There is no plausible X-ray counterpart for source A, which is also
extended approximately in the direction perpendicular to the merger
axis and centered near where the countershock should appear.
Moreover, Figure~\ref{fig:SUMSS} (left panel), an unsharp-masked image
of the Chandra data (the difference of two images convolved with 2D
Gaussian smoothing widths of $\sigma=7.5''$ and $30''$), reveals high
frequency structure, i.e., a sharp edge to the X-ray surface
brightness, at the locations of both radio sources A and C, lending
credence to the idea that we are seeing radio relics similar to those
seen in several nearby clusters, such as Abell 3667
\citep{Rottgering1997}, Abell 3376 \citep{Bagchi2006}, and Abell 1240
\citep{Kempner2001}, that are believed to be powered by merger shocks
\citep[see, e.g.,][]{VanWeeren2011}. \citet{Kempner2004} suggest the
term {\it radio gischt} for these features to distinguish them from
other types of radio sources in clusters, such as AGN relics and radio
phoenix sources.

The monochromatic radio power of source A, assuming it is located at
the redshift of the cluster, is $P_{1.4} = 7\times
10^{25}$~W~Hz$^{-1}$.  Likewise the radio power of source C is
$P_{1.4} = 3\times 10^{25}$~W~Hz$^{-1}$.  The rest-frame frequency of
the 843~MHz observations at the redshift of \J0102\ is 1.576 GHz,
which we multiply by a factor of 1.126 to convert to the monochromatic
power at rest-frame frequency 1.4 GHz, which assumes a spectral index
of $-1$.  The radio power of source A is one of the most intense known
among radio gischt sources.  \citet{Feretti2002} show that the radio
power of gischt relics correlate with the cluster bolometric X-ray
luminosity (albeit with large scatter).  The high radio power and
bolometric X-ray luminosity of \J0102\ put it at the extreme upper end
of this correlation.  For \J0102, an important caveat of our
interpretation of the SUMSS measurements is that the angular
resolution of the radio images is insufficient to identify and remove
radio source contamination, so the values of radio power quoted above
for sources A and C should be considered upper limits to the diffuse
radio emission. Approved observations at the Australia Telescope
Compact Array (ATCA) will provide higher resolution imaging that will
help to remove much of this uncertainty due to unresolved radio source
emission at the site of the relics. If confirmed this will be the most
distant radio relic observed.

\section{Summary and Conclusions}

We present a multi-wavelength analysis of \J0102\ ``El Gordo,'' a new
SZ cluster which appears to be the first reported high-redshift
Bullet-like cluster undergoing a major merger.  We have used a
combination of optical (VLT), X-ray (\chandra), and infrared
(\spitzer) data to investigate the cluster's physical properties and
find that it is an exceptionally massive, X-ray luminous and hot
system, resulting in the highest SZ signal in the ACT survey.  From
the optical spectra of 89 galaxies confirmed as cluster members, we
obtain a spectroscopic redshift ($z=0.87008$) and velocity dispersion
($\sigma_{\rm gal}=1321 \pm 106$~km~s$^{-1}$) for this cluster.  Our
60-ks \chandra\ observation on ACIS-I provides a spectroscopic
temperature of $T_X = 14.5 \pm 1.0$~keV and bolometric luminosity of
$L_{\rm bol}=13.6 \times 10^{45}\, h_{70}^{-2}$ erg s$^{-1}$.  From
mass scaling relations for the velocity dispersion, X-ray $Y_X$, and
SZ distortion, we determine the total mass of the cluster to be
$M_{200a}=(2.16\pm0.32)\times10^{15}\,h_{70}^{-1}M_\odot$, establishing
it as the most massive cluster known at $z>0.6$.

From our warm-phase \spitzer\ IRAC and deep optical imaging, we
measure the total stellar mass and constrain the stellar content of
the cluster to be $<$1\% of the total mass, in broad agreement with
other massive clusters. The optical ($riz$ bands) and IR (\3p6, \4p5)
color-magnitude diagrams for spectroscopically-confirmed members show
a clearly-defined red sequence consistent with passively evolving
galaxies, including a small fraction of actively star-forming
galaxies.  However the BCG is extremely luminous, significantly bluer
than the red sequence, and appears as an E+A+[\ion{O}{2}] galaxy from
the VLT spectra.  Such blue BCGs are normally seen in the centers of
cool core clusters and indeed this BCG sits right within the cool, low
entropy, enhanced-metal-abundance merging core in \J0102 revealed by
the \chandra\ data.

While clusters as massive as \J0102 are quite rare at its redshift,
the cluster does not pose any significant difficulty for the standard
\lcdm\ cosmology provided its mass is in the lower portion of its
statistically allowed mass range.  Our \chandra\ and VLT observations
additionally show that \J0102\ is undergoing a major merger with a
mass ratio of approximately 2 to 1 between its subcomponents.  We find
no analogous high-mass merging systems, with properties broadly
similar to \J0102, within any current large-volume cosmological N-body
simulations (e.g., MICE, Cubep3m).
Thus the high mass and merger ratio at this redshift argues for an
exceptional rarity of \J0102\ within our Universe. We expect that more
detailed analysis of large cosmological simulations and better
understanding of the propagation of the initial Gaussian fluctuations,
particularly at high redshift, will be required to compare with the
predictions of a \lcdm\ model using massive systems like \J0102.

The X-ray surface brightness of our \chandra\ observation reveals a
``wake'' in the hot gas distribution that we attribute to the recent
passage of one cluster through the other. The deprojected gas density
distribution shows that the wake requires a low-density cavity in the
interior of the cluster.  Understanding the evolution of these
features, determining the merger speed and quantifying the rarity of a
merging cluster system with the relative masses and velocities of the
subcomponents of \J0102 will require further simulations and analysis.

Inspection of archival SUMSS radio observations at 843~MHz reveals the
presence of radio sources located at the southeast and northwest edges
of the cluster, which we tentatively identify as radio gischt relics
hosted by \J0102. They lack obvious X-ray counterparts and their
positions relative to the cluster are consistent with the expected
locations of merger-induced shocks. The northwestern radio source is
intense ($P_{1.4} = 7\times 10^{25}$~W~Hz$^{-1}$) and is extended
perpendicular to the merger direction.  If upcoming ATCA observations
confirm the diffuse nature of the radio emission, then \J0102\ will be
the highest redshift cluster with a radio gischt relic.

The next most critical observational step is to obtain strong and weak
lensing mass estimates and maps from the ground and with the {\em
  Hubble Space Telescope}, which we are pursuing. This will allow us
to confirm the presence of an offset between the baryonic and dark
matter distributions, as well as provide a robust mass estimate
unbiased by merger state or cluster astrophysics.  A deeper approved
\chandra\ observation will allow better mapping of the spatial
temperature distribution in the cluster, which may allow
identification of shocks in the gas. It is now possible to use ALMA to
search for these same features directly in the SZ effect using the low
frequency channel (Band 3 covering 84--116 GHz).  Confirmation of the
radio relic is clearly a high priority since these are typically found
in merging clusters; moreover \J0102's high redshift will allow
investigation of their cosmic evolution.  As we have outlined above,
several theoretical steps also need to be pursued in order to fully
understand the many distinctive features of \J0102\ that we have
presented here. We hope that this new exceptional cluster will fuel
more in-depth simulations of cluster mergers and motivate continued
searches for similar high-redshift systems.

\acknowledgements

We are very gratefull to Gustavo Yepes for detailed discussions and
exploring the results of the simulations performed by the Juropa
supercomputer at Juelich.
We would like to thank Ricardo Demarco for helpful discussions on the
spectra of galaxies in the cluster and Larry Rudnick for suggesting we
look at the SUMSS data.
We acknowledge \chandra\ grant number GO1-12008X and \spitzer\
JPL-RSA\#1414522 to Rutgers University.
This work is based in part on observations made with the Spitzer Space
Telescope (PID 70149), which is operated by the Jet Propulsion
Laboratory, California Institute of Technology under a contract with
NASA. Support for this work was provided by NASA through an award
issued by JPL/Caltech.
This work was supported by the U.S. National Science Foundation
through awards AST-0408698 for the ACT project, and PHY-0355328,
AST-0707731 and PIRE-0507768 (award number OISE-0530095). The PIRE
program made possible exchanges between Chile, South Africa, Spain and
the US that enabled this research program. 
Funding was also provided by Princeton University and the University
of Pennsylvania.
Computations were performed on the GPC supercomputer at the SciNet HPC
Consortium. SciNet is funded by: the Canada Foundation for Innovation
under the auspices of Compute Canada; the Government of Ontario;
Ontario Research Fund--Research Excellence; and the University of
Toronto.
This research is partially funded by ``Centro de Astrof\'{\i}sica FONDAP''
15010003, Centro BASAL-CATA and by FONDECYT under proyecto 1085286.
M. Hilton acknowledges financial support from the Leverhulme Trust.

\end{document}